\documentclass[11pt,a4paper]{article}
\pdfoutput=1
\usepackage{jheppub}
\usepackage{rotating}
\usepackage{array}
\usepackage{amsmath,amssymb,amstext,bbold}
\usepackage[normalem]{ulem}
\usepackage{slashed}
\usepackage{booktabs}
\usepackage[pdftex,table,dvipsnames]{xcolor}
\usepackage{units}
\usepackage{multirow}
\usepackage{url}
\usepackage{color}
\usepackage{xspace}
\usepackage{comment}
\preprint{TTK-19-25, P3H-19-019}

\title{Strongly interacting dark sectors in the early Universe and at the LHC through a simplified portal}

\author{Elias Bernreuther,}
\author{Felix Kahlhoefer,}
\author{Michael Kr\"amer}
\author{and Patrick Tunney}

\affiliation{Institute for Theoretical Particle Physics and Cosmology (TTK), RWTH Aachen University, \\ D-52056 Aachen, Germany}

\emailAdd{ebernreuther@physik.rwth-aachen.de}
\emailAdd{kahlhoefer@physik.rwth-aachen.de}
\emailAdd{mkraemer@physik.rwth-aachen.de}
\emailAdd{tunney@physik.rwth-aachen.de}

\abstract{We study the cosmology and LHC phenomenology of a consistent strongly interacting dark sector coupled to Standard Model particles through a generic vector mediator. We lay out the requirements for the model to be cosmologically viable, identify annihilations into dark vector mesons as the dominant dark matter freeze-out process and discuss bounds from direct detection. At the LHC the model predicts dark showers, which can give rise to semi-visible jets or displaced vertices. Existing searches for di-jet resonances and for missing energy mostly probe the parameter regions where prompt decays are expected and constrain our model despite not being optimised for dark showers. We also estimate the sensitivity of dedicated analyses for semi-visible jets and emphasize the complementarity of different search strategies.}


\begin{document}

\maketitle

\flushbottom

\section{Introduction}

A rapidly growing effort is being dedicated to the exploration of dark matter (DM) scenarios where the DM particle does not appear in isolation but as part of a richer dark sector, which may in particular feature new strong interactions~\cite{Strassler:2006im,Hochberg:2014kqa,Hansen:2015yaa,Kribs:2016cew,Choi:2018iit,Kribs:2018oad,Choi:2019zeb,Brax:2019koq}. Such dark sectors provide a variety of mechanisms to set the DM relic abundance~\cite{Hochberg:2014dra,Dienes:2016vei,Choi:2017mkk,Heikinheimo:2018esa,Beauchesne:2018myj} and lead to novel signatures at collider and fixed-target experiments~\cite{Hochberg:2015vrg,Englert:2016knz,Berlin:2018tvf,Renner:2018fhh,Hochberg:2018rjs,Kribs:2018ilo}. Furthermore, dark sectors with significant self-interactions may explain astrophysical small scale structure observations that are in tension with predictions of collisionless cold DM simulations~\cite{Tulin:2017ara}.

The phenomenology of the model depends decisively on the internal properties of the dark sector, in particular the number of dark quark flavours and their respective charge assignments. These properties determine which mesons are stable and which mesons can decay into Standard Model (SM) particles, as well as the processes that can contribute to the freeze-out of the dark sector. Based on these considerations, one obtains rather strong constraints on the structure of the dark sector if cosmological constraints are to be satisfied and the observed DM relic abundance is to be reproduced. In this paper we focus on a QCD-like $SU(3)$ dark sector, in which the dark quarks form various bound states, in particular dark pseudoscalars $\pi$ and vector mesons $\rho$. 
We identify the case of 2 dark quark flavours as particularly interesting since this guarantees that all dark pions are stable, avoiding strong constraints from the decay of dark sector particles in the late and early Universe.\footnote{As we explain later, the late decay of dark pions is strongly constrained by nucleosynthesis and recombination, whereas early decays would keep the dark pions in thermal equilibrium with the Standard Model, leading to an unacceptably suppressed relic density.} The dark pion stability allows us to study the phenomenology of strongly interacting dark sectors at ground based experiments while simultaneously being consistent with all cosmological constraints. In our set up we find that the DM relic density is set by the annihilation process $\pi\pi \to \rho\rho$, which is kinematically forbidden in the present Universe~\cite{DAgnolo:2015ujb}.

At the same time, there is substantial freedom in the form of the portal interaction that determines how the dark sector as a whole couples to the SM. The main effect of such interactions is that they induce decays of dark $\rho^0$ mesons into SM particles. Relatively weak interactions are sufficient for the $\rho^0$ decays to proceed sufficiently fast to keep the dark sector in thermal equilibrium with the SM during freeze-out, while somewhat stronger interactions allow for the dark sector to be probed with direct detection experiments and at the LHC. Rho mesons that decay promptly on collider scales as well as long-lived mesons are a generic possibility of these models, leading to a wide range of novel collider signatures, such as semi-visible jets~\cite{Cohen:2015toa,Cohen:2017pzm,Pierce:2017taw,Beauchesne:2017yhh} and emerging jets~\cite{Schwaller:2015gea,Renner:2018fhh}.

In the present work we therefore adopt a hybrid approach, in which the dark sector is modelled in as much detail as possible (given the inherent limitations from non-perturbative physics), while the interactions with the SM are described using a simplified model similar to the ones commonly used in the context of LHC DM searches~\cite{Abdallah:2014hon,Abdallah:2015ter}. Specifically, we consider a spin-1 mediator $Z'$ with vector couplings to both SM and dark quarks and no other interactions. For this coupling structure direct detection constraints are known to be quite strong if the DM mass is sufficiently above the GeV scale~\cite{Chala:2015ama,Boveia:2016mrp}. However, the typical mass scale of the dark sector is largely unconstrained and can easily be of the order of a few GeV, such that direct detection constraints are weakened. If furthermore the $Z'$ mediator has a mass at the TeV scale, direct detection constraints are additionally suppressed, while LHC constraints become relevant, leading to an interesting interplay between the two different search strategies.

While simplified models of strongly interacting dark sectors have been proposed previously in the literature (see e.g.\ Ref.~\cite{Daci:2015hca}), our approach differs in that the dark sector is constructed in a way that guarantees a consistent cosmology, independent of the details of the portal interaction. Conversely, our approach differs from most models of Strongly Interacting Massive Particles (SIMPs) with a $Z'$ or dark photon mediator in the literature~\cite{Lee:2015gsa,Berlin:2018tvf} in that we allow for a more general coupling structure, in which the interactions of the $Z'$ are dominated by its direct couplings to SM fermions (rather than interactions induced by mixing). 

We find that large parts of the interesting parameter space are presently unconstrained by direct detection experiments and by LHC searches for di-jet resonances or missing transverse energy. However, the fact that a typical dark shower contains both stable and unstable dark mesons means that existing LHC searches are not optimised for the characteristic signatures of strongly interacting dark sectors. We propose dedicated searches for dark showers, which offer great potential to explore the model in more detail, and point out the relevance of searches for displaced vertices~\cite{Alimena:2019zri}. At the same time, significant efforts are being made to develop new direct detection strategies for low-mass DM, which will substantially improve the sensitivity to light dark sectors~\cite{Battaglieri:2017aum}.

This paper is structured as follows. In section~\ref{sec:model} we present the model that we consider and derive the corresponding chiral Lagrangian. A particular emphasis is placed on the discussion of meson stability. A broad overview of the phenomenology of the model is then given in section~\ref{sec:pheno}. We calculate the lifetime of the unstable particles, the relic density and direct detection constraints. Finally, section~\ref{sec:LHC} takes a closer look at various LHC searches. We first consider existing constraints from searches with visible and invisible final states and then discuss the potential of dedicated searches for the specific signatures of our model. Our conclusions are presented in section~\ref{sec:conclusions}.

\section{Dark sector model set-up}
\label{sec:model}

We consider a dark sector consisting of $N_f$ flavours of dark quarks $q_\mathrm{d} = \left(q_{\mathrm{d},i}\right)$ with $i=1 \dots N_f$, which are in the fundamental representation of a dark $SU(N_\mathrm{d})$ gauge group. The corresponding Lagrangian is given by
\begin{align}
\label{eq:lagrangian_massless}
\mathcal{L}=-\frac{1}{4}F_{\mu\nu}^a F^{\mu\nu a} + \overline{q}_{\mathrm{d}} i \slashed{D} q_{\mathrm{d}} - \overline{q}_{\mathrm{d}} M_q q_{\mathrm{d}}\; ,
\end{align}
where $M_q$ denotes the dark quark mass matrix.
In order to have a theory that resembles QCD, we assume $N_\mathrm{d} = 3$. For reasons that will become clear below, we furthermore restrict ourselves to the case $N_f = 2$. 

The dark sector described by eq.~(\ref{eq:lagrangian_massless}) is completely secluded from the SM. Such secluded dark sectors can have a viable cosmology, for example in models with asymmetric reheating~\cite{Adshead:2016xxj,Hardy:2017wkr}, but make few testable predictions. We therefore focus on the case that there is an additional interaction between the two sectors, which establishes thermal equilibrium in the early Universe and allows for the exchange of entropy. We assume that these interactions arise from an additional $U(1)^\prime$ gauge group under which both dark quarks and SM quarks are charged. The $U(1)^\prime$ is broken such that the corresponding $Z^\prime$ gauge boson acquires a mass $m_{Z^\prime}$. The two dark quarks are taken to have opposite charges under the $U(1)^\prime$ such that the interactions with the $Z'$ can be written as
\begin{align}
\mathcal{L} \supset - e_\mathrm{d} Z^\prime_\mu \left(\overline{q}_{\mathrm{d},1}\gamma^\mu q_{\mathrm{d},1}- \overline{q}_{\mathrm{d},2}\gamma^\mu q_{\mathrm{d},2}\right)  = - e_\mathrm{d} Z^\prime_\mu \, \overline{q}_\mathrm{d} Q \gamma^\mu q_\mathrm{d}  \; ,
\end{align}
where $e_\mathrm{d}$ denotes the product of the $U(1)^\prime$ gauge coupling and the charge of the dark quarks and $Q=\mathrm{diag}(1,-1)$ is the dark quark charge matrix. The assignment of the $U(1)^\prime$ charge $Q$ is of relevance to the stability of dark mesons, as discussed in detail below. On the SM side, we consider a universal coupling of the $Z^\prime$ to all SM quarks:
\begin{equation}
\mathcal{L} \supset -Z^\prime_\mu \; g_q \sum_{q_\mathrm{SM}} \overline{q}_\mathrm{SM}\gamma^\mu q_\mathrm{SM} \; .
\end{equation}
Couplings to leptons, as well as mixing between the $Z'$ and SM gauge bosons, are assumed to be sufficiently suppressed to be irrelevant for phenomenology. Our set-up hence resembles the simplified model of a vector mediator that is frequently used for the interpretation of DM searches at the LHC~\cite{Abdallah:2014hon,Abdallah:2015ter}.

At some scale $\Lambda_\mathrm{d}$ the dark sector confines
and the dark quarks form bounds states in the form of dark mesons and dark baryons. In the present work we will focus on the dark mesons, assuming that the dark baryons are sufficiently heavy that they are cosmologically irrelevant and do not contribute to the present-day DM density. More specifically, we will be interested in the pseudoscalar mesons $\pi$ and the vector mesons $\rho$.\footnote{Throughout this work $\pi$ and $\rho$ always refer to dark mesons, not to their SM counterparts.} The  former can be understood as the Goldstone bosons associated with chiral symmetry breaking $SU(N_f)_L \times SU(N_f)_R \to SU(N_f)_V$. It follows that the number of pions equals the number of broken generators $T^a$, with $a = 1\dots N_f^2-1$. We use the normalization $\mathrm{Tr}(T^aT^b)=\delta^{ab}/2$. For $N_f=2$, the pion matrix thus reads
\begin{align}
\label{eq:pionmatrix}
\pi \equiv \pi^a T^a = \frac{1}{\sqrt{2}}
\begin{pmatrix}
\frac{\pi^0}{\sqrt{2}} & \pi^+\\
\pi^- & -\frac{\pi^0}{\sqrt{2}}
\end{pmatrix} \; ,
\end{align}
where $\pi^0$, $\pi^+$, $\pi^-$ denote the $U(1)^\prime$ charge eigenstates, i.e.\ the $\pi^{\pm}$ have charge $\pm2e_\mathrm{d}$ and the $\pi^0$ are uncharged. As we will discuss in more detail below, in our set-up all three pions are stable and hence they contribute equally to the DM relic abundance.

Below the confinement scale, the interactions of the pseudoscalar mesons can be described by a chiral effective field theory (EFT), written in terms of
\begin{align}
U = \exp\left(2i\pi/f_\pi\right) \; ,
\end{align}
where $f_\pi$ denotes the dark pion decay constant. 
For example, the mass term for the pseudoscalars is given by~\cite{Berlin:2018tvf}
\begin{align}
\label{eq:chiraleft_mass}
\mathcal{L}_\mathrm{mass} = \frac{f_\pi^2 B_0}{2}\mathrm{Tr}\left(M_qU^\dagger\right) + \text{h.c.} \; ,
\end{align}
where $B_0$ is a dimensionless constant. If both dark quark masses are the same, i.e.\ $M_q = \text{diag}(m_q, m_q)$, one finds
\begin{align}
\label{eq:chiraleft_mass2}
\frac{f_\pi^2 B_0}{2}\mathrm{Tr}\left(M_qU^\dagger\right) + \text{h.c.} \; = \; m_\pi^2\mathrm{Tr}\left(\pi^2\right) + \mathcal{O}\left(\frac{\pi^4}{f_\pi^2}\right)
\end{align}
with $m_\pi^2 = 2B_0m_q$. The detailed expressions for the interactions of dark pions with each other as well as the interactions between dark pions and the $Z'$ gauge boson are provided in appendix~\ref{app:lagrangian}.

The vector mesons can be parametrised as~\cite{Berlin:2018tvf,klingl1996}\footnote{In general, $V_\mu$ also contains the vector $\omega$, and $\pi$ contains the pseudoscalar $\eta$. However, we assume that these particles are sufficiently heavy that they play no important role in the phenomenology of the model and can therefore be omitted in this study.}
\begin{align}
\label{eq:vectormesonmatrix}
V_\mu = V_\mu^aT^a = \frac{1}{\sqrt{2}}
\begin{pmatrix}
\frac{\rho^0_\mu}{\sqrt{2}} & \rho^+_\mu \\
\rho^-_\mu & -\frac{\rho^0_\mu}{\sqrt{2}}
\end{pmatrix} \; .
\end{align}
It is furthermore helpful to define the vector meson field strength
\begin{align}
\label{eq:vectormeson_fieldstrength}
V_{\mu\nu} = \partial_\mu V_\nu - \partial_\nu V_\mu - ig\left[V_\mu,V_\nu\right] \; ,
\end{align}
where $g$ is the pion-vector-meson coupling strength. The so-called KSRF relation~\cite{ks1966,fr1966}, which relates properties of the $\rho$ meson to the pion decay constant, implies~\cite{Berlin:2018tvf}
\begin{align}
\label{eq:ksfr}
g \approx \frac{m_\rho}{\sqrt{2}f_\pi} \; ,
\end{align}
which we use to express $f_\pi$ in terms of $m_\rho$ and $g$. The detailed expressions for the interactions between the pseudoscalar and vector mesons can again be found in appendix~\ref{app:lagrangian}.

The interactions between the $Z^\prime$ and the charged vector mesons arise from the term
\begin{align}
\label{eq:zpv_mixing}
\mathcal{L} \supset -\frac{e_\mathrm{d}}{g_{Z^\prime V}} Z^\prime_{\mu\nu} \mathrm{Tr}\left(QV^{\mu\nu}\right) \; ,
\end{align}
which induces in particular the $\rho^+\rho^-Z'$ vertex~\cite{klingl1996}. For this vertex to have the correct normalisation consistent with the charge $\pm 2 e_\mathrm{d}$ of $\rho^\pm$ requires that $g_{Z^\prime V} = g$, i.e.\ the pion-vector-meson coupling strength defined above. Furthermore, the term \eqref{eq:zpv_mixing} gives rise to mixing between the $Z'$ and the $\rho^0$. This mixing is of central importance for the phenomenology of our model, as it induces small couplings between the $\rho^0$ and SM quarks, which render the $\rho^0$ unstable (see appendix~\ref{app:lagrangian}). A detailed calculation of the resulting $\rho^0$ lifetime will be provided in section~\ref{sec:lifetime}.

To summarise, in the perturbative regime our model can be completely characterised by the five parameters $m_q$, $\Lambda_\mathrm{d}$, $m_{Z^\prime}$, $e_\mathrm{d}$ and $g_q$. In the confinement regime, the first two parameters are replaced by the three effective parameters $m_{\pi}$, $m_\rho$ and $g$. An overview of all particles in the dark sector and the corresponding parameters is given in Tab.~\ref{tab:parameters}.

\begin{table}[tb]
	\begin{center}
		\begin{tabular}{l|l|l}
			\toprule
			Sector & Particles & Parameters \\
			\midrule
			$SU(3)$  & $q_{\mathrm{d}, i}$~($i=1,2$), $A_\mathrm{d}$~(dark gluon) &  $m_q$, $\Lambda_\mathrm{d}$ \\
			Chiral EFT & $\pi^\pm$, $\pi^0$, $\rho^\pm$, $\rho^0$ & $m_{\pi}$, $m_\rho$, $g$ \\
			$U(1)^\prime$ & $Z^\prime$ & $m_{Z^\prime}$, $e_\mathrm{d}$, $g_q$ \\
			\bottomrule
		\end{tabular}
	\end{center}
	\caption{Overview of independent parameters in the model.}
	\label{tab:parameters}
\end{table}

Before exploring the phenomenology of our model in detail, let us briefly discuss how our model differs from similar scenarios with three flavours, as considered for example in~\cite{hochberg2014}. First of all, for $N_f = 3$, the Chiral EFT Lagrangian includes the Wess-Zumino-Witten (WZW) term~\cite{witten1983,wess1971}
\begin{align}
\label{eq:wzwterm}
\frac{2N_\mathrm{d}}{15\pi^2f_\pi^5} \epsilon^{\mu\nu\rho\sigma}\mathrm{Tr}\left(\pi\partial_\mu\pi\partial_\nu\pi\partial_\rho\pi\partial_\sigma\pi\right) \; ,
\end{align}
which induces the $3\pi \to 2\pi$ annihilations crucial to the SIMP mechanism~\cite{hochberg2014}. For $N_f=2$ this term vanishes due to its antisymmetry under pion exchange, so that we need to consider alternative mechanisms for obtaining the DM relic abundance (see section~\ref{sec:relic}). Further interaction terms involving $\pi$ and $\rho$ arise when the WZW term \eqref{eq:wzwterm} is gauged~\cite{klingl1996}. 
However, for $N_f=2$ all anomalous $\pi$-$\rho$ interactions vanish, because $\mathrm{Tr}\left(\sigma^a\{\sigma^b,\sigma^c\}\right)=0$ and therefore the $SU(2)$ is anomaly-free.

The most important difference however concerns the pion stability. While the charged dark pions are guaranteed to be stable if there are no lighter states carrying $U(1)'$ charge, the $\pi^0$ can in principle decay into $q\overline{q}$ or $q\overline{q}q\overline{q}$. Such decays result for example from the triangle anomaly, which gives rise to a $\pi^0Z^\prime Z^\prime$ vertex in complete analogy to the $\pi^0_\mathrm{SM}\gamma\gamma$ vertex in the SM. Even if the anomaly vanishes (i.e.\ if the dark quark charge matrix satisfies $Q^2\propto \mathbb{1}$), neutral pions can still decay via higher-order terms of the WZW type in the chiral Lagrangian~\cite{bijnens1988}. 
Such decays turn out to be extremely dangerous for the viability of the model. If $\pi^0$ decays are fast, they will keep the pions in equilibrium with the SM and lead to an exponential suppression of the DM relic abundance. Slow decays, on the other hand, are subject to strong constraints from nucleosynthesis and recombination.

To evade these constraints, one can impose a dark $G$-parity~\cite{Berlin:2018tvf}
\begin{align}
\label{eq:dark_gparity}
G_\mathrm{d} = C' \times \mathbb{Z}_2 \times U_q \; ,
\end{align}
where $C'$ is the $U(1)^\prime$ charge conjugation operator, and we have introduced a $\mathbb{Z}_2$ transformation that takes $Z^\prime \to -Z^\prime$, as well as an $SU(N_f)$ transformation $U_q$ that satisfies
\begin{align}
\label{eq:uq_chargematrix}
U_q^\dagger \; Q \; U_q = -Q^T \; .
\end{align}
In combination with the requirement $Q^2\propto \mathbb{1}$, eq.~\eqref{eq:uq_chargematrix} can only be satisfied if the number $N_f$ of dark quarks is even. Specifically, for $N_f = 2$ and $Q=\mathrm{diag}(1,-1)$ one finds
\begin{align}
U_q =
\begin{pmatrix}
0 & 1 \\
-1 & 0
\end{pmatrix} \: ,
\end{align}
such that $\pi^0$ is indeed odd under $G_\mathrm{d}$. For $N_f = 3$, on the other hand, it is not possible to define an analogous $G$-parity and hence $\pi^0$ decays cannot be forbidden.

\section{General phenomenology}
\label{sec:pheno}

For the discussion above there was no need to specify the mass scale of the dark sector or the mass of the $Z'$ mediator. From now on we will be more specific and consider a GeV-scale dark sector that interacts with the SM via a TeV-scale $Z'$. As discussed in detail below, GeV-scale dark sectors provide various interesting mechanisms to produce the thermal relic density, and are less constrained by direct detection experiments than heavier DM particles. The phenomenology of dark sectors with light $Z'$ mediators has been explored in the literature, see e.g.~\cite{Berlin:2018tvf}. Here we focus on heavy TeV-scale mediators which lead to new avenues for dark sector searches at the LHC.

Since the $Z'$ is heavy compared to the confinement scale of the dark sector, we can calculate its production and decay in terms of free (SM and dark) quarks~-- this will be the topic of section~\ref{sec:LHC}. At low energies and in the present Universe, however, the appropriate degrees of freedom are the pseudoscalar and vector mesons that appear in the confinement phase. For the reasons outlined above, all dark pions are exactly stable in the model we consider and can therefore potentially account for the DM in the Universe. In the present section, we will study the mechanisms that determine the relic abundance of dark pions, as well as constraints from direct detection experiments. Before doing so, we however need to consider the properties of the rho mesons, which turn out to be of central importance for the phenomenology of the model.

\subsection{$\rho^0$ lifetime}
\label{sec:lifetime}

In the presence of the interaction term in eq.~(\ref{eq:zpv_mixing}) the $\rho^0$ meson mixes with the $Z'$ boson. This mixing gives rise to interactions between the $\rho^0$ and SM quarks, which to leading order in $m_\rho / m_{Z'}$ can be written as
\begin{equation}
 \mathcal{L}_\text{EFT} \supset \frac{2 \, e_\mathrm{d} \, g_q}{g} \frac{m_\rho^2}{m_{Z'}^2} \rho^{0\mu} \sum_{q_{\mathrm{SM}}}^{} \overline{q}_\mathrm{SM}\gamma_\mu q_\mathrm{SM} \; .
\end{equation}
For large $Z'$ masses the induced couplings can be extremely small and hence the $\rho^0$ can potentially be rather long-lived. Calculating the $\rho^0$ lifetime for masses in the GeV range is a notoriously difficult problem. For $m_\rho \gtrsim 2\,\mathrm{GeV}$ and away from any spin-1 QCD resonances, one can employ the perturbative spectator model to estimate $\Gamma_{\rho^0} \approx \sum_{q_\text{SM}} \Gamma(\rho^0 \to q_\text{SM} \overline{q}_\text{SM})$, where
\begin{align}
\Gamma\left(\rho^0\to q_\text{SM} \overline{q}_\text{SM}\right) =
 \frac{1}{\pi}\frac{g_q^2e_\mathrm{d}^2}{g^2}\,m_\rho\left(\frac{m_\rho}{m_{Z'}}\right)^4\left(1-4\frac{m_{q_\text{SM}}^2}{m_\rho^2}\right)^{1/2} \left(1+2\frac{m_{q_\text{SM}}^2}{m_\rho^2}\right)
\end{align}
and the sum includes all quarks that are kinematically allowed, i.e.\ with $m_{q_\text{SM}} < m_\rho / 2$. 

For $m_\rho < 2\,\mathrm{GeV}$, decays of the $\rho^0$ are more accurately described by calculating the mixing with QCD resonances, in particular with the SM $\rho$ meson. However, since the $Z'$ in our model, and hence the $\rho^0$, couples to all quarks with equal strength, tree-level interactions with SM mesons are absent. Decays into mesons can therefore only proceed via baryon loops, so that we expect the decay width of the $\rho^0$ to become very small in this mass range. In the present work, we will not consider this problem further and focus on $m_\rho \gtrsim 2$~GeV. Finally, we assume that, contrary to QCD, $m_\rho < 2 \, m_\pi$ such that decays into the dark sector are not possible kinematically. Such a small mass difference can arise if the masses of the dark quarks are comparable to the confinement scale, such that the explicit breaking of chiral symmetry is stronger than for QCD.

For concreteness let us introduce a benchmark scenario that we will study in detail in the following. We set $e_\mathrm{d} = 0.4$, $g = 1$, $m_\rho = 5 \, \mathrm{GeV}$ and $m_{Z'} = 1\,\mathrm{TeV}$. We then find
\begin{equation}
 \Gamma_{\rho^0} \approx 0.6\,\mathrm{eV} \times g_q^2 \; ,
\end{equation}
which corresponds to
\begin{equation}
 c \tau_{\rho^0} \approx 3.2 \, \mathrm{mm} \times \left(\frac{g_q}{0.01}\right)^{-2} \; .
 \label{eq:decay_length}
\end{equation}
As we shall discuss in section \ref{sec:LHC}, small couplings $g_q$ lead to displaced $\rho^0$ decays and thus interesting LHC phenomenology.

The $\rho^\pm$ can in principle decay into $\pi^\pm q\overline{q}$ via an off-shell $Z'$ coupled to $\pi$ and $\rho$ through an anomalous vertex of the gauged WZW type. However, for $m_{Z'}\gg m_\rho$, we find this channel to be extremely suppressed by the three-body nature of the decay as well as powers of momentum in the vertex factor. In addition, if $m_\rho-m_\pi\lesssim 2$~GeV, the appropriate final states would again be SM mesons rather than quarks. As discussed above, however, tree-level couplings of the $Z'$ to SM mesons are absent. We can therefore treat $\rho^\pm$ as effectively stable during dark sector freeze-out. As we will see below, the abundance of $\rho^\pm$ is exponentially suppressed relative to the dark pion abundance, so that these particles do not contribute to the DM and their late-time decays do not lead to any observable effects.

Apart from the interesting phenomenological implications to be discussed below, the interactions between the $\rho^0$ and SM quarks play an important role in the cosmological evolution. Indeed, the (inverse) decays of the $\rho^0$ are the primary way in which the dark sector can maintain thermal equilibrium with the SM bath. 
For these decays to be efficient, one requires that
\begin{align}
\Gamma_{\rho^0} \gtrsim H(T) \; ,
\label{eq:therm}
\end{align}
where $H \sim 14.4 \, T^2 / M_\mathrm{P}$ is the Hubble rate before the QCD phase transition. For our benchmark point this corresponds to
\begin{equation}
 g_q \gtrsim 4 \cdot 10^{-5} \times \left(\frac{T}{1\,\mathrm{GeV}}\right) \; .
\end{equation}
If this condition is satisfied, the temperature of the $\rho^0$ is equal to the SM temperature and the number density $n_{\rho^0}$ is given by an equilibrium distribution, $n_{\rho^0} = n_\rho^\text{eq}$. The strong interactions between the $\rho^0$ and the $\rho^\pm$ will then ensure that the same also holds for the charged rho mesons. Initially, interactions between the dark pions and rhos will also keep the dark pions in equilibrium. DM freeze-out happens when these interactions become inefficient and the dark pions decouple from the rho mesons, which will be discussed next.

\subsection{Relic density from forbidden annihilations}
\label{sec:relic}

\begin{figure}[t]
	\centering
	\includegraphics[width=0.3\columnwidth]{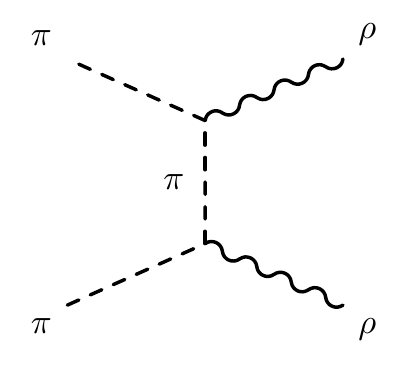}
	\includegraphics[width=0.3\columnwidth]{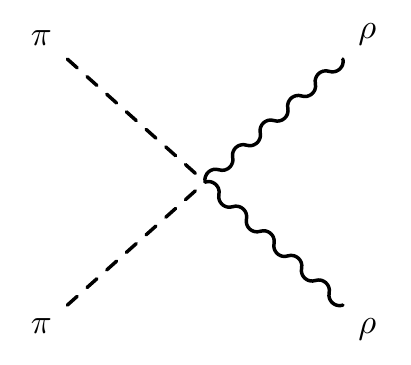}
	\includegraphics[width=0.3\columnwidth]{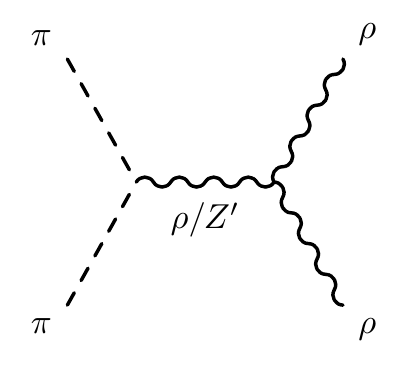}
	\caption{Diagrams contributing to forbidden dark pion annihilations.} \label{fig:forbidden_annihilation}
\end{figure}

The main process that keeps dark pions in thermal equilibrium with the dark rhos (and hence with the SM) is the pair annihilation $\pi \pi \to \rho \rho$ (see figure~\ref{fig:forbidden_annihilation}). This process is fully efficient as long as the temperature is large compared to $m_\rho - m_\pi$, but becomes exponentially suppressed for smaller temperatures~\cite{DAgnolo:2015ujb,Berlin:2018tvf,Li:2019ulz}. 
Provided that the dark $\rho$ mesons are always in thermal equilibrium with the SM, the Boltzmann equation for the dark pion number density $n_\pi$ reads
\begin{align}
\label{eq:boltzmann_pi}
\dot{n}_\pi + 3Hn_\pi = -\langle\sigma_{\pi\pi\to\rho\rho}v\rangle n_\pi^2 + \langle\sigma_{\rho\rho\to\pi\pi}v\rangle (n_\rho^\mathrm{eq})^2 \; .
\end{align}
Since the right-hand side has to vanish if both the dark pions and the dark rhos are in equilibrium, the cross section for forbidden annihilations can be expressed as
\begin{align}
\label{eq:pitorho_xsec}
\langle\sigma_{\pi\pi\to\rho\rho}v\rangle = \langle\sigma_{\rho\rho\to\pi\pi}v\rangle \left(\frac{n_\rho^\mathrm{eq}}{n_\pi^\mathrm{eq}}\right)^2 \; .
\end{align}
Here, $\langle\sigma_{\rho\rho\to\pi\pi}v\rangle$ is unsuppressed at low temperatures and scales proportional to $g^4/m_\pi^2$. The exponential suppression arises from the factor
\begin{equation}
 \left( \frac{n_\rho^\mathrm{eq}}{n_\pi^\mathrm{eq}}\right)^2 \sim \exp(-2 \Delta x)
\end{equation}
with $x = m_\pi / T$ and $\Delta=(m_\rho-m_\pi)/m_\pi$. 
Inserting eq.~\eqref{eq:pitorho_xsec}, the Boltzmann equation \eqref{eq:boltzmann_pi} takes the form
\begin{align}
\label{eq:boltzmann_pi_2}
\dot{n}_\pi + 3Hn_\pi = -\langle\sigma_{\rho\rho\to\pi\pi}v\rangle \left(\frac{n_\rho^\mathrm{eq}}{n_\pi^\mathrm{eq}}\right)^2 \left(n_\pi^2-(n_\pi^\mathrm{eq})^2\right) \; .
\end{align}
From \eqref{eq:boltzmann_pi_2} we can read off that $\pi\pi\to\rho\rho$ interactions decouple when
\begin{align}
\label{eq:pion_decoupling}
\langle\sigma_{\rho\rho\to\pi\pi}v\rangle \frac{(n_\rho^\mathrm{eq})^2}{n_\pi^\mathrm{eq}} \approx H.
\end{align}
For example, for the benchmark point above and $m_\pi = 4\,\mathrm{GeV}$, such that $\Delta = 0.25$, one finds that decoupling happens for $x \approx 26$. The resulting abundance of dark pions is found to be close to the observed value $\Omega_\text{DM} h^2 = 0.12$~\cite{Aghanim:2018eyx}. Due to the exponential sensitivity of $\Omega_\text{DM}$ to the mass splitting $\Delta$, it is always possible to reproduce the observed value by varying $m_\pi$ (or alternatively $m_\rho$) slightly.

We compute the relic density using \textsc{MicrOMEGAs} 5.0.6~\cite{Belanger:2018mqt}, which allows for up to two dark species in its freeze-out calculation. Conversions between $\pi^\pm$ and $\pi^0$ remain efficient until long after annihilations of dark pions into dark rho mesons have frozen out. Therefore, we can treat all dark pions as one species in \textsc{MicrOMEGAs}. The $\rho^\pm$ are assigned as a second dark species, whose abundance is strongly suppressed at low temperatures.\footnote{We note that the $\rho^\pm$ are not necessarily stable and may decay with a very long lifetime, $\tau_{\rho^\pm} \gtrsim 1 \,\mathrm{s}$. Hence, for sufficiently small couplings they may induce interesting signatures in cosmology and indirect detection experiments even if their abundance is tiny. A closer investigation of these effects is beyond the scope of the present work.}
Furthermore, in the freeze-out scenario considered here, $\rho^0$ is in equilibrium with the SM throughout dark-pion freeze-out, and can hence be treated like an SM particle by \textsc{MicrOMEGAs}.

Let us briefly revisit our assumption that the dark rho mesons are in thermal equilibrium with the SM during dark pion freeze-out. 
In order for the dark rhos to remain in equilibrium, (inverse) decays should be at least as efficient as the conversion between dark rhos and pions until the point when the dark pions freeze out. This requires that
\begin{align}
n_\rho^\mathrm{eq}\, \Gamma_{\rho^0} > (n_\pi^\mathrm{eq})^2 \langle\sigma_{\pi\pi\to\rho\rho}v\rangle > n_\pi^\mathrm{eq} \, H \; ,
\end{align}
before dark pion freeze-out, and hence
\begin{align}
\frac{n_\rho^\mathrm{eq}}{n_\pi^\mathrm{eq}} \Gamma_{\rho^0} \gtrsim H \; .
\label{eq:rhodecays}
\end{align}
Hence, requiring that rho decays not be a limiting factor for pion freeze-out yields the lower bound \eqref{eq:rhodecays} on $\Gamma_{\rho^0}$, which is more stringent than the simple requirement of thermalisation in eq.~\eqref{eq:therm}.

\subsection{Constraints from direct detection experiments}

At low energies dark rho exchange induces an effective coupling of $\pi^\pm$ to SM nucleons $N = p,n$, given by
\begin{align}
\mathcal{O}_N &= \frac{6 \, e_\mathrm{d} \, g_q}{m_{Z^\prime}^2} \left[\pi^+\left(\partial_\mu\pi^-\right)-\left(\partial_\mu\pi^+\right)\pi^-\right] \, \overline{N}\gamma^\mu N \; ,
\end{align}
which turns out to depend on $m_{Z'}$ rather than $m_\rho$ because of the way in which the interactions arise from $\rho$--$Z'$ mixing (see appendix~\ref{app:lagrangian}). This effective interaction gives rise to spin-independent scattering with cross section given by
\begin{align}
\sigma_N^\mathrm{SI} = \frac{36 \, e_\mathrm{d}^2 \, g_q^2 \, \mu_{\pi N}^2}{\pi \, m_{Z^\prime}^4} \; ,
\end{align}
where $\mu_{\pi N} = m_\pi \, m_N / (m_\pi + m_N)$ is the reduced mass.

Since we have assumed universal quark couplings, the DM-nucleus cross section receives a coherent enhancement proportional to the square of the mass number $A$. We can therefore directly compare our model predictions for $\sigma_N^\mathrm{SI}$ to published exclusion limits and obtain a bound on the effective coupling $e_\mathrm{d} \, g_q / m_{Z'}^2$ as a function of $m_\pi$. However, it is important to account for the fact that neutral pions do not couple to SM quarks at tree-level. Thus, for the purpose of direct detection the effective local DM density is reduced by a factor 2/3, which can be captured by an appropriate rescaling of published exclusion limits.

For the mass range $1\,\mathrm{GeV} \lesssim m_\pi \lesssim 10\,\mathrm{GeV}$ that we will be interested in, relevant constraints arise from a number of different direct detection experiments, namely CRESST-III~\cite{Abdelhameed:2019hmk}, CDMSLite~\cite{Agnese:2017jvy}, DarkSide-50~\cite{Agnes:2018ves}, PICO-60~\cite{Amole:2019fdf}, PandaX~\cite{Cui:2017nnn} and XENON1T~\cite{Aprile:2018dbl}. Rather than simply considering each experimental result separately, we use \textsc{DDCalc} 2.0~\cite{Athron:2018hpc} to perform a statistical combination of all experimental results. However, as explained in detail in appendix~\ref{app:darkside}, we do not include the DarkSide-50 analysis, which relies on an overly optimistic extrapolation of the ionisation yield to low energies. In addition, we separately consider sensitivity projections for the SuperCDMS experiment~\cite{Agnese:2016cpb}, which should provide substantial improvements for small DM masses.

\begin{figure}[t]
\centering
\includegraphics[width=0.6\columnwidth]{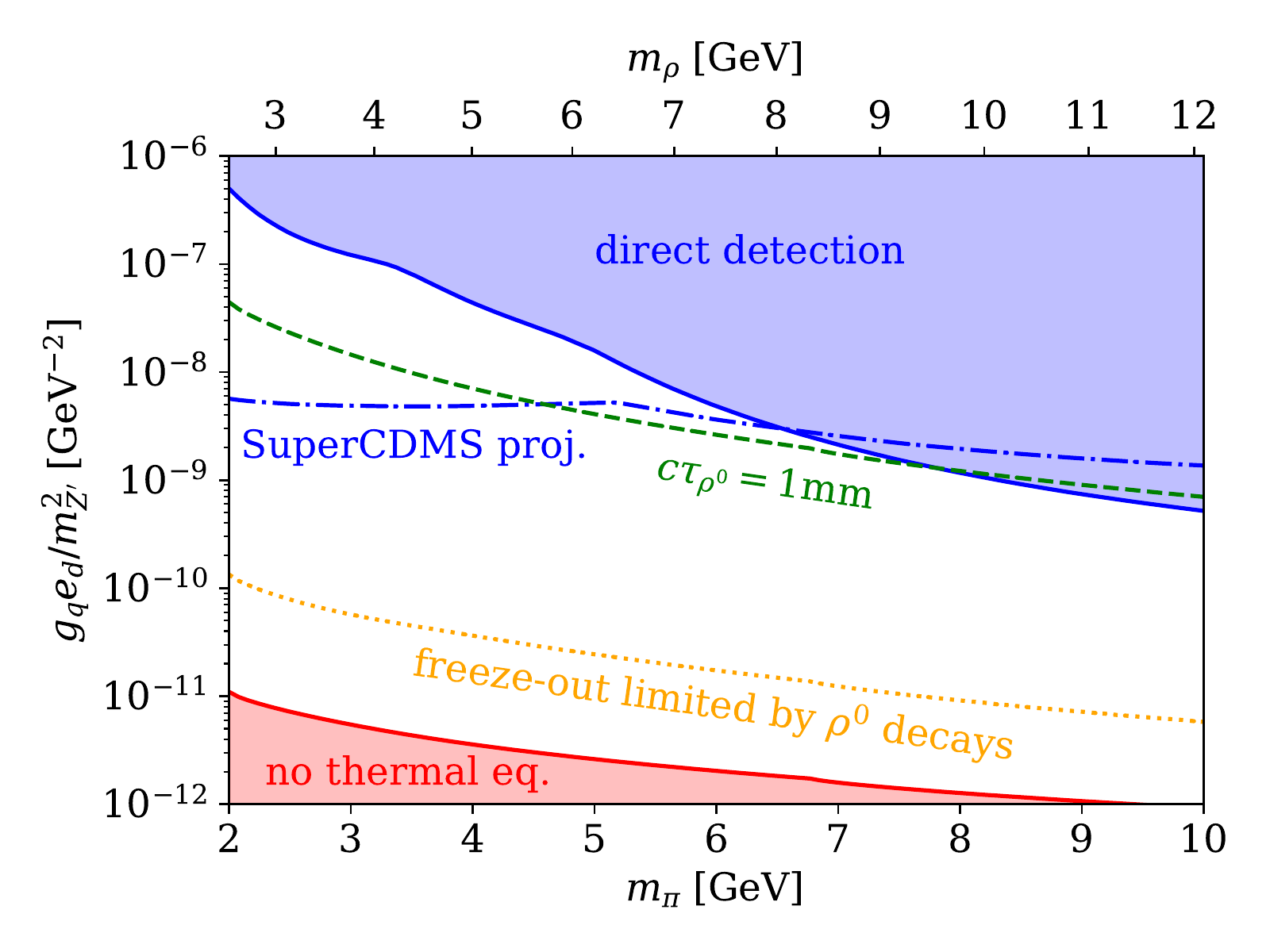}
\caption{Constraints on the dark pion mass $m_\pi$ and effective coupling $g_q e_\mathrm{d} / m_{Z'}^2$ from a combined analysis of different direct detection experiments (see text for details). The second $x$-axis at the top indicates the values of $m_\rho$ for which the observed DM relic abundance can be reproduced for a $\pi$-$\rho$ coupling of $g = 1$. The relic density calculation requires the $\rho^0$ mesons to be in equilibrium with the SM until the end of freeze-out, which is satisfied for the parameter region above the orange dashed line. Below the green dashed line, the $\rho^0$ decay length may be macroscopic. \label{fig:directdetection}}
\end{figure}

The resulting constraints on the parameter space are shown in figure~\ref{fig:directdetection}. While the direct detection constraints depend only on the parameters that are being varied explicitly, we can include additional information in the figure by fixing the $\pi$-$\rho$ coupling $g$. Doing so enables us to calculate $m_\rho$ as a function of $m_\pi$ from the relic density requirement, as indicated by the second $x$-axis at the top. For given $m_\rho$ we can then determine the $\rho^0$ decay length and indicate the parameter region where one can expect LHC signatures with displaced vertices. Furthermore, we show the parameter region where the requirement of thermal equilibrium during dark pion freeze-out is violated (see eq.~\eqref{eq:therm}), as well as the parameter region where $\rho^0$ decays are inefficient (see eq.~\eqref{eq:rhodecays}).

\subsection{Astrophysical constraints}
Let us finally consider potential constraints from astrophysical observations. Indirect detection experiments typically give strong constraints on models of thermal DM with a particle mass below 10 GeV. In our model, however, the dominant annihilation channel for dark pions is $\pi \pi \to \rho \rho$. Since $m_\rho > m_\pi$, this channel is only kinematically open in the early Universe and becomes strongly suppressed for small relative velocities. Thus, there are no relevant indirect detection constraints for our model.

One of the attractive features of strongly interacting dark sectors is their potential to resolve the so-called small-scale crisis of $\Lambda$CDM with DM self-interactions. In our model, the self-scattering cross section for $\pi^+ \pi^- \to \pi^+ \pi^-$ divided by the dark pion mass is given by
\begin{equation}
 \frac{\sigma_\text{self}}{m_\pi} = \frac{m_\pi}{4 \pi \, f_\pi^4} \sim 10^{-3} \, \mathrm{cm^2 / g} \left(\frac{m_\pi}{1\,\mathrm{GeV}}\right)^{-3} \left(\frac{g}{\sqrt{4\pi}}\right)^4 \; . 
\end{equation}
Similar results hold for the scattering channels involving neutral pions.

Hence, for GeV-scale dark pion masses the self-scattering cross section is too small to induce observable effects in small astrophysical systems, which would require $\sigma_\text{self} / m_\pi \gtrsim 1\,\mathrm{cm^2 / g}$. For the same reason, the parameter space that we consider in this work easily satisfies all constraints on the self-interaction cross section from merging galaxy clusters. 

\section{LHC constraints}
\label{sec:LHC}

The LHC phenomenology of our model is dominated by the on-shell production of a $Z'$ boson and its subsequent decay. Since the mass of the $Z'$ is assumed to be large compared to the confinement scale of the dark sector, the partial decay widths can be calculated in terms of free quarks:
\begin{align}
\label{eq:width_zptoqq}
 \Gamma(Z' \to q_\text{SM} \overline{q}_\text{SM}) & = \sum_{q_\text{SM}} \frac{g_q^2}{4 \pi} m_{Z'}\left(1 + 2 \frac{m_{q_\mathrm{SM}}^2}{m_{Z'}^2} \right) \sqrt{1 - \frac{4 \, m_{q_\mathrm{SM}}^2}{m_{Z'}^2}} \; , \\
\label{eq:width_zptoqdqd}
 \Gamma(Z' \to q_\mathrm{d} \overline{q}_\mathrm{d}) & = \frac{e_\mathrm{d}^2}{2 \pi} m_{Z'} \left(1 + 2 \frac{m_q^2}{m_{Z'}^2} \right) \sqrt{1 - \frac{4 \, m_q^2}{m_{Z'}^2}} \; ,
\end{align}
where the sum runs over all SM quarks and $m_q$ denotes the corresponding quark mass. In the absence of any other decay modes, the branching ratio into dark sector states is then given by
\begin{equation}
\label{eq:br_zptoqdqd}
 \text{BR}(Z' \to q_\mathrm{d} \overline{q}_\mathrm{d}) = \frac{\Gamma(Z' \to q_\mathrm{d} \overline{q}_\mathrm{d})}{\Gamma(Z' \to q_\mathrm{d} \overline{q}_\mathrm{d}) + \Gamma(Z' \to q_\text{SM} \overline{q}_\text{SM})} \; .
\end{equation}
It is worth noting that $\Gamma(Z' \to q_\mathrm{d} \overline{q}_\mathrm{d})$ is proportional to $N_f \times N_\mathrm{d}$ and hence the branching ratio into dark sector states can be sizeable without the need for a large hierarchy of couplings between $e_\mathrm{d}$ and $g_q$. For example, for $e_\mathrm{d} = 0.4$ and $g_q = 0.1$, the branching ratio into dark sector states is found to be $84\%$. 

The case of the $Z'$ decaying into SM quarks is constrained by LHC searches for di-jet resonances. A variety of search strategies have been developed to search for such resonances in different mass ranges. The strongest bound for high-mass resonances stems from a recent ATLAS search based on an integrated luminosity of $\mathcal{L} = 139 \, \mathrm{fb^{-1}}$~\cite{ATLAS:2019bov}. A compilation of different constraints on low-mass resonances was recently presented in Ref.~\cite{Ellis:2018xal}. We apply these constraints to our model by rescaling them with the appropriate branching ratios, which is a good approximation because the total width of the $Z'$ is sufficiently small.

\begin{figure}[t]
\centering
\includegraphics[width=0.5\textwidth]{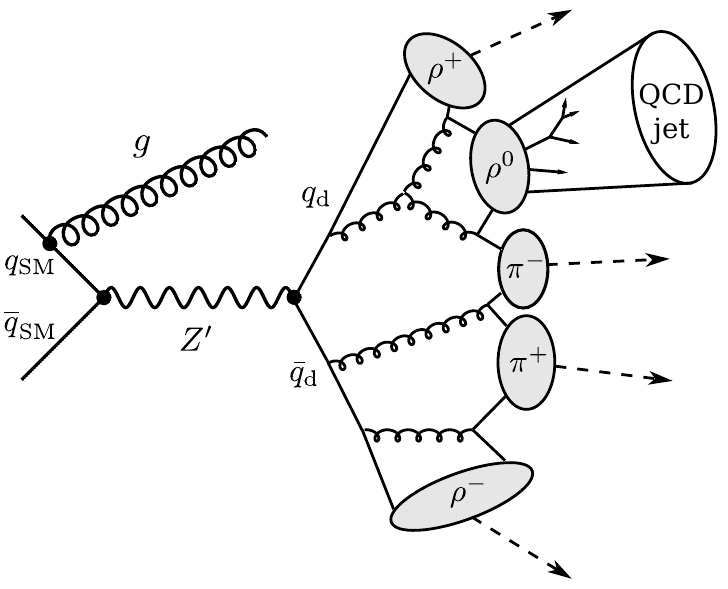}
\caption{Schematic illustration of a dark shower from the decay of a $Z'$ produced in association with a gluon.}\label{fig:darkshower}
\end{figure}

If the $Z'$ decays into dark quarks, fragmentation and hadronisation proceeds within the dark sector (see figure~\ref{fig:darkshower}). We simulate these dark showers using the Hidden Valley model in \textsc{Pythia}~\cite{Carloni:2010tw,Carloni:2011kk}, which calculates the final yield and distribution of dark mesons. The number of dark mesons produced within each shower depends on the initial energy of the dark quark and varies from event to event. On average, 10--12 dark mesons are produced from a dark quark with an energy of 1 TeV, but events with more than 20 mesons per shower are not uncommon.

Accordingly, we find the typical boost factor for dark mesons to be of the order of $\overline{\gamma} \equiv \overline{E}_\rho / m_{\rho} \approx 10$, where $\overline{E}_\rho$ denotes the average energy of $\rho$ mesons. Because of this boost factor, the actual decay length of the $\rho^0$ mesons in the laboratory frame, $\gamma c \tau$, is substantially larger than the estimate given in eq.~(\ref{eq:decay_length}). In the following, we will assume that for $\overline{\gamma} c \tau < 1 \,\mathrm{mm}$ the rho meson decays can be treated as prompt, such that conventional experimental strategies apply. However, since both the boost factor and the actual distance travelled before the decay are subject to large fluctuations, displaced vertices may be observable even for smaller decay lengths.

The average relative multiplicity of the different mesons depends on their respective number of degrees of freedom. Spin-1 $\rho$ mesons are three times as abundant as spin-0 $\pi$ mesons and charged $\rho^\pm$ and $\pi^\pm$ mesons are twice as abundant as their neutral partners. It follows that we expect on average $25\%$ of a dark shower to consist of $\rho^0$ mesons, which subsequently decay into SM hadrons, while the remaining $75\%$ are stable mesons that escape from the detector unseen. A dark shower will hence lead to a semi-visible jet~\cite{Cohen:2015toa,Cohen:2017pzm} with an average fraction of invisible energy of $r_\text{inv} = 0.75$.

Such semi-visible jets give rise to a range of interesting experimental signatures. If the $Z'$ is produced in isolation, i.e.\ without additional energetic SM particles from initial state radiation (ISR), the two semi-visible jets will be back-to-back. Defining the minimum angular separation in the azimuthal plane between the missing energy vector $\slashed{E}_T$ and any of the leading jets
\begin{equation}
 \label{eq:deltaphi}
 \Delta \phi = \underset{j}{\text{min}} \, \Delta \phi(j, \slashed{E}_T) \; ,
\end{equation}
such events are expected to have small $\Delta \phi$, as the missing energy is aligned with one of the dark showers. Ordinary ''mono-jet`` searches (i.e.\ searches for energetic jets in association with missing energy) will reject such events because of prohibitive QCD backgrounds from misreconstructed jets~\cite{Aaboud:2017phn,Sirunyan:2017jix}. Traditional searches for di-jet resonances are also expected to be insensitive to these kinds of events, since the visible jets only carry a fraction of the energy of the dark shower and hence their invariant mass does not peak at the mass of the $Z'$.

However, given the relatively large value of $r_\text{inv}$, there is a non-negligible probability for a dark shower to remain entirely invisible. In this case, the $Z'$ decay would lead to a missing energy vector pointing away from the visible jets, such that $\Delta \phi \approx \pi$ and the event selection cuts applied in mono-jet searches can be satisfied. While the missing energy that can be obtained in this way is limited to $\slashed{E}_T < m_{Z'} / 2$, larger amounts of missing energy are possible if the $Z'$ recoils against an ISR jet (see figure~\ref{fig:darkshower}). In this case, also events with two partially visible jets can contribute, as the missing energy vector will in general not be aligned with any of the jets.

\begin{figure}[t]
	\centering
	\includegraphics[width=0.495\columnwidth]{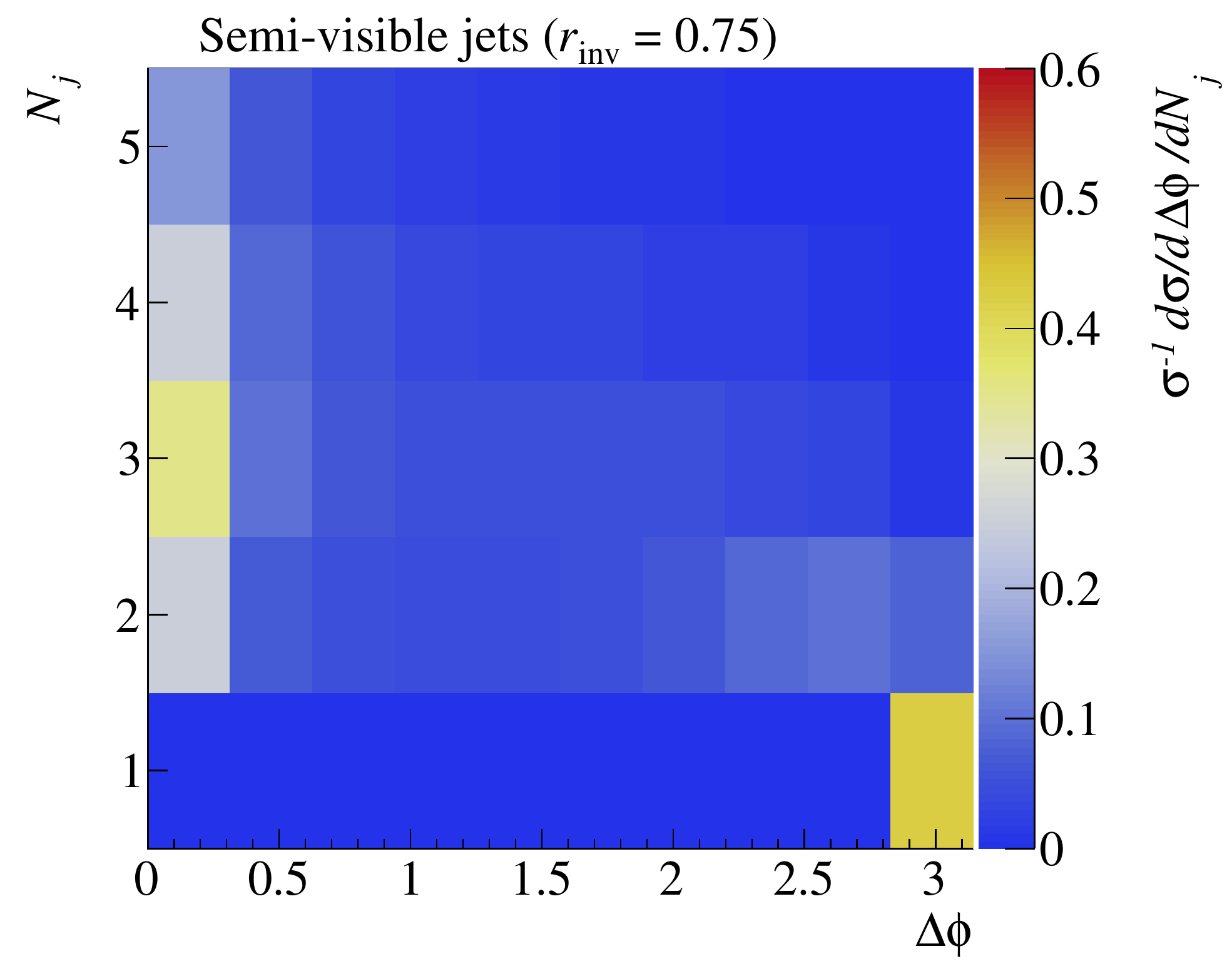}
	\includegraphics[width=0.495\columnwidth]{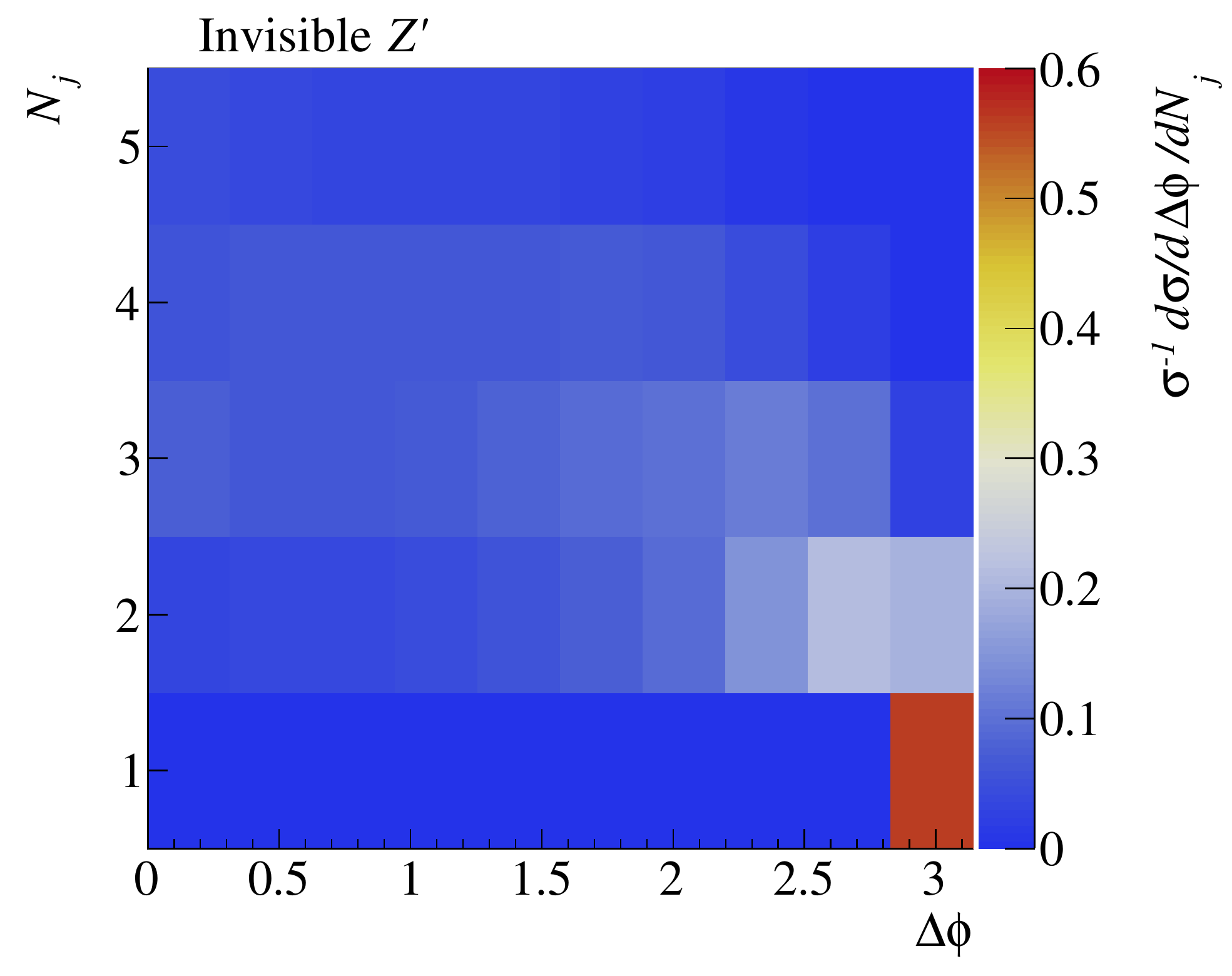}
	\caption{Double differential cross section with respect to $\Delta\phi$ and $N_j$ for the dark sector model (left) and a fully invisible $Z'$ (right) with $m_{Z'} = 1\,\mathrm{TeV}$. For both panels, the typical mono-jet requirement $\slashed{E}_T>250\,\mathrm{GeV}$ has been applied. \label{fig:deltaphi_nj}}
\end{figure}

These considerations are illustrated in figure~\ref{fig:deltaphi_nj}, which shows the double differential cross section with respect to $\Delta\phi$ and $N_j$ after imposing the requirement $\slashed{E}_T>250\,\mathrm{GeV}$. In the left panel, we consider the case of dark showers with $r_\text{inv} = 0.75$, whereas the right panel corresponds to fully invisible $Z'$ decays, such that jets can only arise from ISR. In the first case, one can clearly see the two contributions discussed above, where either $N_j = 1$ and $\Delta \phi \approx \pi$ or $N_j > 1$ and $\Delta \phi \approx 0$. For the case of fully invisible decays, on the other hand, the second contribution is absent.

In the following, we will first consider existing constraints on the parameter space from searches for di-jet resonances and searches for jets in association with missing transverse energy. Afterwards, we discuss a novel search strategy that targets semi-visible jets more specifically and estimate the corresponding sensitivity. As we will see, all these searches probe the parameter regions where the $\rho^0$ is short-lived and decays promptly on collider scales. Nevertheless, even for smaller couplings the $Z'$ production cross section can be sizeable, leading to events with displaced vertices, which would look similar to the emerging jet signature discussed in Ref.~\cite{Schwaller:2015gea}. A detailed analysis of such signatures will be left to future work.

\subsection{Constraints from missing energy searches}
\label{sec:met_searches}

Mono-jet signatures can arise in our model if the dark showers remain mostly invisible and recoil against an ISR jet, or if some part of the dark showers produces an energetic jet while the rest remains invisible. Moreover, since mono-jet searches typically allow for more than one jet, events with multiple jets from ISR or the dark shower can also contribute. On the other hand, events where both dark showers are partially visible can have similar kinematics as squark or gluino pair production. In both cases, events consist of two hemispheres with a jet and missing energy in each hemisphere. Hence, supersymmetry searches for jets and missing energy can be sensitive to dark shower production.

\paragraph*{Simulation details.}

Having implemented our model with the \textsc{FeynRules} package~\cite{FeynRules}, we generate parton-level events at leading order for the dark quark production process $p p \to q_\text{d}\overline{q}_\text{d}$ with \textsc{MadGraph5\_aMC{@}NLO}~\cite{Madgraph} using the \textsc{NN23LO1} PDF set~\cite{NNPDF}. We perform MLM matching of samples with up to one hard jet setting $\text{xqcut}=100$~GeV. The $Z'$ width is calculated self-consistently by \textsc{MadGraph} at each parameter point. The parton-level events are then showered and hadronised, both in QCD and in the dark sector, with \textsc{Pythia  8}~\cite{Pythia8}. 

The simulation of the dark shower and of dark meson production relies on \textsc{Pythia}'s Hidden Valley module~\cite{Carloni:2010tw}. The hidden valley meson states pivUp, pivDn, pivDiag, rhovUp, rhovDn and rhovDiag provided by \textsc{Pythia8} are equivalent to the dark mesons $\pi^+$, $\pi^-$, $\pi^0$, $\rho^+$, $\rho^-$ and $\rho^0$, respectively. Out of these, only $\rho^0$ (rhoDiag) decays into SM quarks while the others are stable. We adjust the Hidden Valley variable probVector appropriately to implement the expected ratio of invisible mesons $r_\text{inv}=0.75$. Furthermore, we turn on the running of the dark coupling $\alpha_\text{d}$ determined by the confinement scale $\Lambda_\text{d}$.

We scan over a range of $Z'$ masses between $500$~GeV and $5000$~GeV, generating $10^5$ events for each $Z'$ mass. The other relevant parameters are set to the benchmark values $m_{q}=500$~MeV, $m_\pi=4$~GeV, and $m_\rho=\Lambda_\mathrm{d}=5$~GeV. Larger meson masses would reduce the average meson multiplicity, while different values of $m_q$ and $\Lambda_\mathrm{d}$ would not significantly change our results.

\paragraph*{Recasting and analysis details.}

We recast existing experimental analyses with \textsc{CheckMATE 2}~\cite{CheckMATE2,CheckMATE}, and \textsc{MadAnalysis 5}~\cite{MadAnalysis5,MadAnalysis5_2} in conjunction with its Public Analysis Database (PAD)~\cite{MadAnalysis5_pad}. Both these codes first pass the hadron-level events from \textsc{Pythia8} to \textsc{Delphes3}~\cite{Delphes3}, which simulates the appropriate detector for the respective analysis. \textsc{Delphes3} internally calls the \textsc{FastJet}~\cite{FastJet,FastJet_2} package for jet clustering. Subsequently, the analysis cuts are applied to the signal events and $95\%$-exclusion limits $S^{95}$ on the number of signal events are calculated based on the number of observed background events. To incorporate the signal uncertainty without the time-consuming computation of CLs values~\cite{Read:2002hq}, \textsc{CheckMATE} calculates the ratio $r$ defined as
\begin{align}
r = \frac{S - 1.64 \Delta S}{S^{95}} \; ,
\end{align}
where $S$ denotes the predicted number of signal events and $\Delta S$ the corresponding Monte Carlo uncertainty.

First we consider the most recent ATLAS search for mono-jet events at $\sqrt{s}=13$~TeV and with an integrated luminosity of $36.1\,\mathrm{fb^{-1}}$~\cite{ATLAS:2017dnw,Aaboud:2017phn} as implemented in \textsc{CheckMATE}. Jets are clustered using the anti-$k_T$ algorithm~\cite{Cacciari:2008gp} with $R=0.4$. The search requires $\slashed{E}_T>250$~GeV, and $p_T>250$~GeV for the leading jet. Events are allowed to have at most four jets with $p_T>30$~GeV. For the angular distance between the missing energy and the leading jets the search requires that $\Delta\phi > 0.4$, with $\Delta\phi$ as in eq.~(\ref{eq:deltaphi}).
A range of inclusive and exclusive signal regions are defined in terms of the amount of missing transverse energy.

Among SUSY searches available for recasting we find the most recent CMS squark and gluino search~\cite{Sirunyan:2017cwe} to have the highest sensitivity to dark shower signal in the $Z'$ mass range we consider. This search uses $35.9\,\mathrm{fb^{-1}}$ of data at $\sqrt{s}=13$~TeV and is implemented in the \textsc{MadAnalysis5} PAD~\cite{CMS_SUS_16_033_MA5}. In contrast to the mono-jet analysis, here at least 2 jets are required. Events need to fulfil $\slashed{E}_T>300$~GeV and $H_T>300$~GeV, where $H_T$ denotes the scalar sum of transverse momenta of all jets. Moreover, $\Delta\phi(j,\slashed{E}_T)>0.5$ is imposed on the two leading jets, and $\Delta\phi(j,\slashed{E}_T)>0.3$ on the third and fourth jet. Each signal region is defined by the combination of an $\slashed{E}_T$ range and an $H_T$ range.

For each $Z'$ mass, we conservatively approximate the exclusion bound from a given search by the limit from the most sensitive signal region, based on observed event numbers.
Since $\Gamma_{Z'}/m_{Z'}$, as given in eqs.~\eqref{eq:width_zptoqq} and \eqref{eq:width_zptoqdqd}, is well below $10\%$ in the relevant parameter space, we translate bounds on the number of signal events $S$ into bounds on couplings and masses using the narrow width approximation, i.e.
\begin{align}
\sigma\left(p p \to q_\text{d}\overline{q}_\text{d}\right) \approx \sigma\left(p p \to Z'\right) \times \text{BR}(Z' \to q_\mathrm{d} \overline{q}_\mathrm{d}) \; ,
\end{align}
with the branching ratio given in \eqref{eq:br_zptoqdqd}.

\begin{figure}[t]
	\centering
	\includegraphics[width=0.495\columnwidth]{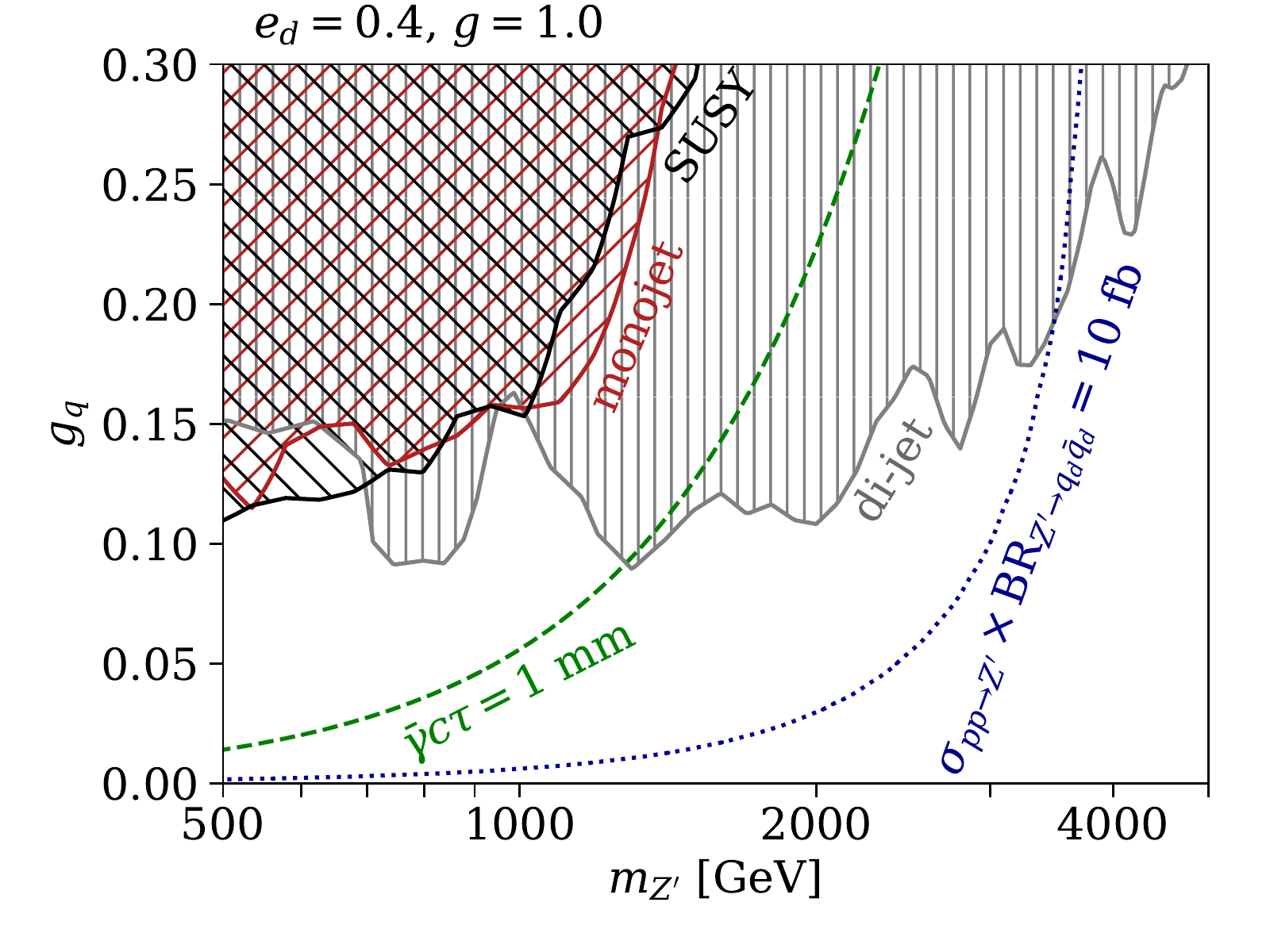}
	\includegraphics[width=0.495\columnwidth]{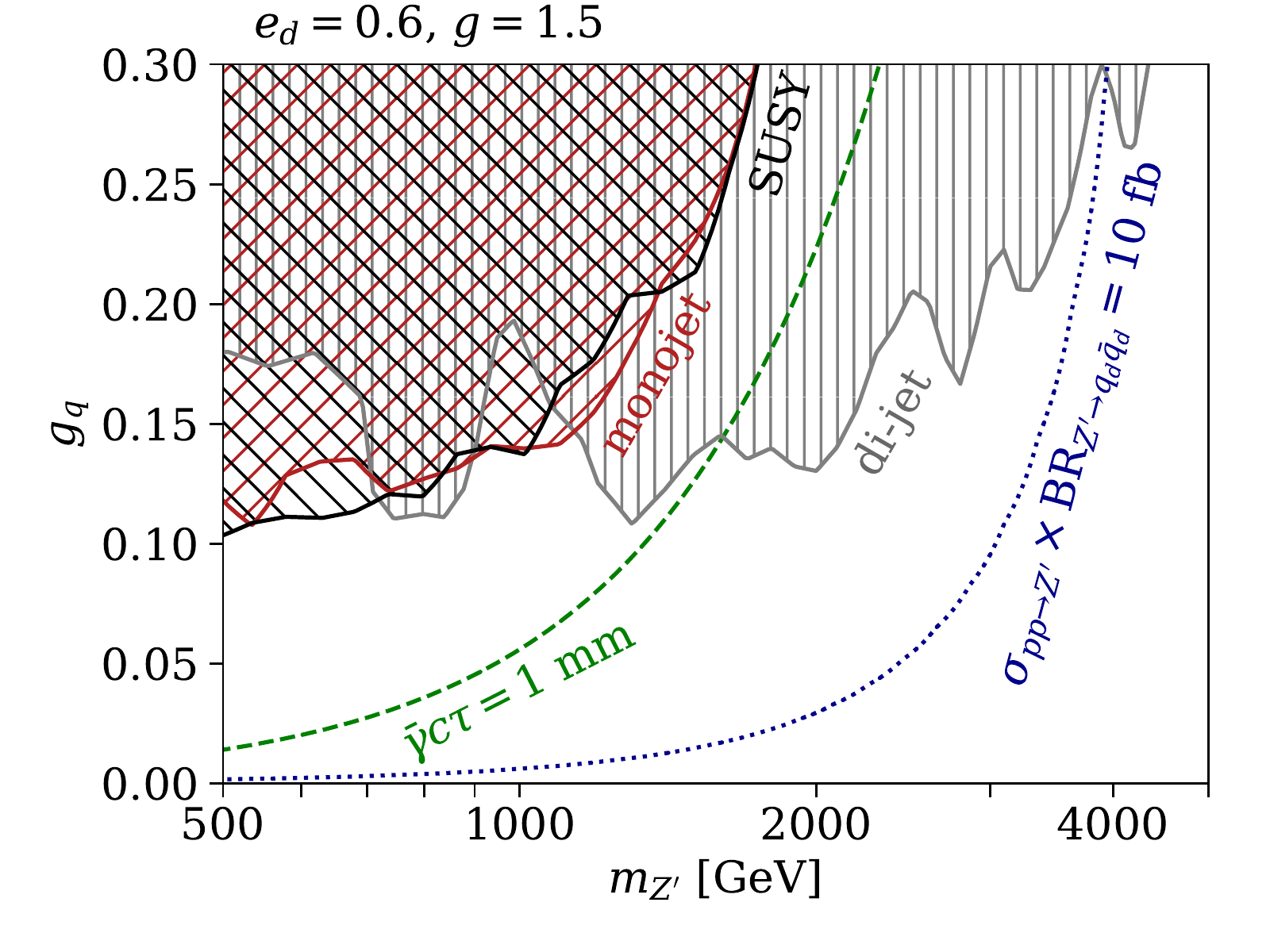}
	\caption{Summary of existing LHC constraints on the dark sector model considered in this work. Above the green dashed line it is a good approximation to treat $\rho^0$ decays as prompt, whereas below the line one expects displaced decays.\label{fig:lhc_limits_etmiss}}
\end{figure}

We present our results in figure~\ref{fig:lhc_limits_etmiss}, which shows the parameter regions excluded by searches for di-jet resonances and searches for missing energy in the $g_q$-$m_{Z'}$ parameter plane for different values of $e_\mathrm{d}$. Larger values of $e_\mathrm{d}$ suppress the branching ratio of the $Z'$ into SM quarks and therefore the impact of di-jet constraints while enhancing the sensitivity of searches for missing energy. For $m_{Z'} \lesssim 2\,\mathrm{TeV}$ the combined constraints imply $g_q \lesssim 0.1$, while for $m_{Z'} > 4\,\mathrm{TeV}$ values of $g_q$ as large as 0.3 are allowed.

We also indicate the combination of parameters corresponding to $\overline{\gamma} c \tau_{\rho^0} = 1\,\mathrm{mm}$ for $\overline{\gamma} \approx 10$. As discussed above, $\rho^0$ decays can be treated as prompt above this line, so that the analyses discussed above can be safely applied. The search for di-jet resonances, on the other hand, does not involve any dark mesons and is therefore not affected by the $\rho^0$ lifetime. Below the green dashed line, one would generically expect displaced vertices. We emphasise that the number of such events can potentially be quite large. For concreteness, we show the combinations of $g_q$ and $m_{Z'}$ that correspond to $\sigma_{pp\to Z'} = 10\,\mathrm{fb}$, which would correspond to more than $10^3$ such events having been produced to date at ATLAS and CMS.

To conclude this discussion, we emphasise that the analyses considered above have not been optimised to search for dark showers. The mono-jet search, for example, is intended to search for $Z'$ bosons that decay fully invisibly. In this case, signal events are expected to have large $\Delta \phi$ (see figure~\ref{fig:deltaphi_nj}), so that a cut on $\Delta \phi$ substantially improves the signal-to-background ratio. In our case the situation is quite different, as events with $N_j > 1$ and small $\Delta \phi$ contribute substantially to the signal cross section. This makes it important to reassess whether the requirement $\Delta \phi > 0.4$ is strictly necessary to suppress backgrounds. Conversely, it may be interesting to reject events with $N_j > 1$ and $\Delta \phi >  2$, which are more likely to arise from background than from our model.

\subsection{Prospective searches for semi-visible jets}

Having reinterpreted existing generic searches for jets and missing energy, let us now discuss dedicated searches for the dark showers produced in the decay of the $Z'$ boson. If the dark showers were completely visible, one would obtain a peak in the total invariant mass of the resulting jets, which would be centred at the $Z'$ mass and could be used to distinguish signal from background. However, for semi-visible jets the peak in $M_{jj}^2=(p_{j_1}+p_{j_2})^2$ is washed out to a considerable extent by the fact that the fraction of visible energy in the jets differs from event to event. 

We therefore consider an analysis proposed in Ref.~\cite{Cohen:2015toa,Cohen:2017pzm}, which is based on the transverse jet mass
\begin{align}
M_T = \left(M_{jj}^2+2\left(\sqrt{M_{jj}^2+p_{Tjj}^2}\slashed{E}_T-\vec{p}_{Tjj}\cdot\vec{\slashed{E}}_T\right)\right)^{1/2} \; ,
\end{align}
where $\vec{p}_{Tjj}$ is the vector sum of the $\vec{p}_T$ of the two jets.
It was shown in Ref.~\cite{Cohen:2017pzm} that this variable maintains the ability to distinguish signal from background up to rather large values of $r_\text{inv}$ and can be competitive with ordinary missing energy searches for $r_\text{inv} \approx 0.75$.

To apply this analysis to our model we pass the events generated by the tool chain described in section~\ref{sec:met_searches} to \textsc{Delphes3} with CMS settings and use the anti-$k_T$ algorithm with $R=0.5$ for jet clustering. We then apply the same cuts as in Ref.~\cite{Cohen:2017pzm}. At the preselection level, these require $\slashed{E}_T>200$~GeV and at least two jets with $p_T>100$~GeV and $\lvert\eta\rvert<2.4$. Subsequently, jets are re-clustered with the Cambridge-Aachen algorithm and $R=1.1$. We then compute the transverse mass
of the two leading re-clustered jets and require $\slashed{E}_T/M_T > 0.15$. Moreover, we impose $\lvert\eta_{j_1}-\eta_{j_2}\rvert<1.1$. Electrons with $p_T>10$~GeV and $\lvert\eta\rvert<2.4$ as well as muons with $p_T>20$~GeV and $\lvert\eta\rvert<2.4$ are vetoed. Finally, the analysis imposes an angular separation cut that is inverted compared to the mono-jet case and requires $\Delta\phi<0.4$ between the missing energy vector and at least one jet.

\begin{figure}[t]
	\centering
	\includegraphics[width=0.57\columnwidth]{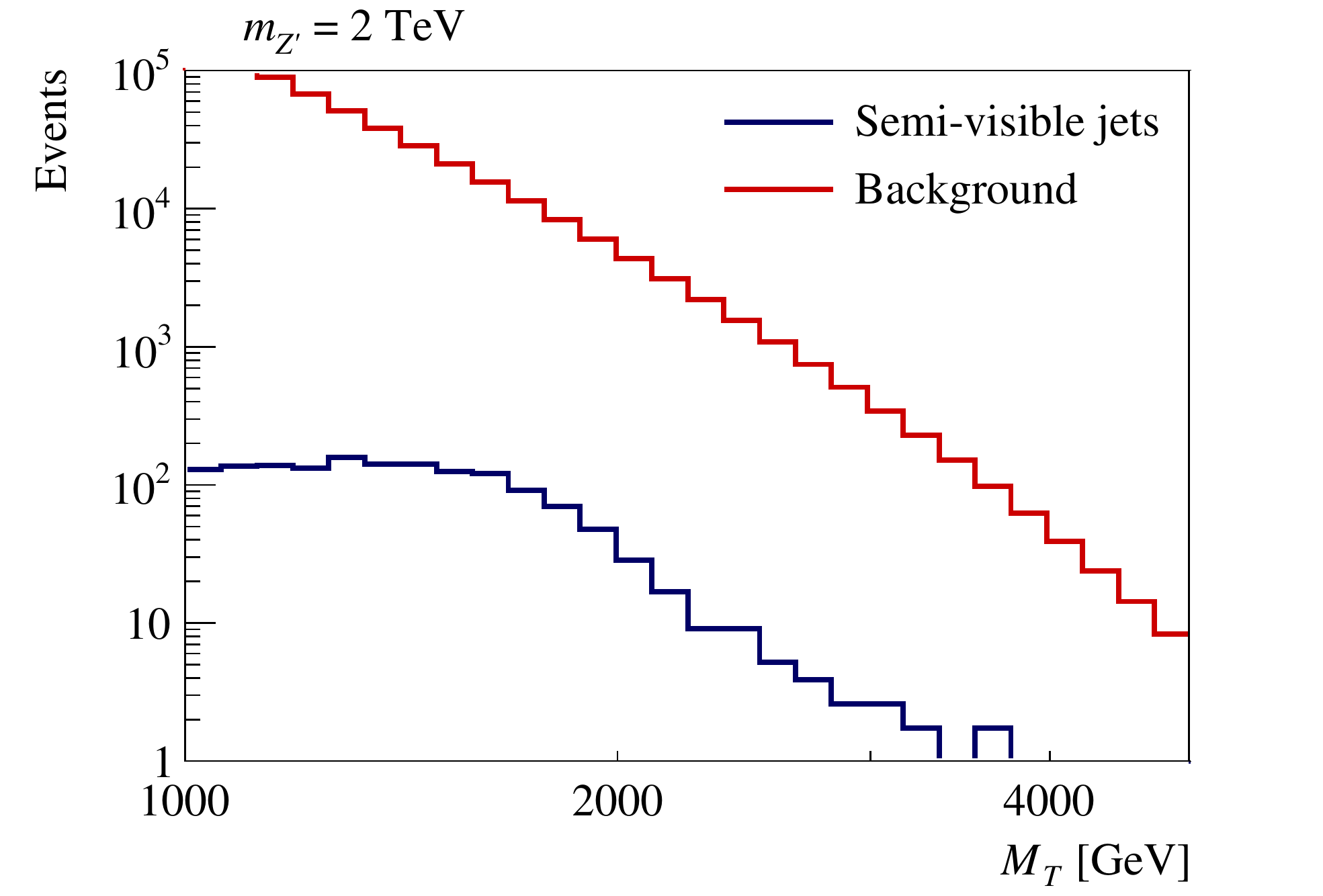}
	\caption{Differential cross section with respect to the transverse jet mass $M_T$ for semi-visible jets obtained from the dark showers produced by the decays of a $Z'$ boson with $m_{Z'} = 2\,\mathrm{TeV}$ and for the SM backgrounds simulated in Ref.~\cite{Cohen:2017pzm}. \label{fig:MTspectrum}}
\end{figure}

We show in figure~\ref{fig:MTspectrum} the $M_T$ spectrum expected for a $Z'$ boson with $m_{Z'} = 2 \,\mathrm{TeV}$ decaying into two dark showers with $r_\text{inv} = 0.75$. We have furthermore set $e_d = 0.6$ and $g_q = 0.1$, compatible with current di-jet constraints (see figure~\ref{fig:lhc_limits_etmiss}). For comparison, we also show the background estimate obtained in Ref.~\cite{Cohen:2017pzm} rescaled to $\mathcal{L} = 300\,\mathrm{fb^{-1}}$. One can clearly see the difference in shape between signal and background arising from the fact that $M_T \leq m_{Z'}$ for the decay of an on-shell $Z'$.

\begin{figure}[t]
	\centering
	\includegraphics[width=0.495\columnwidth]{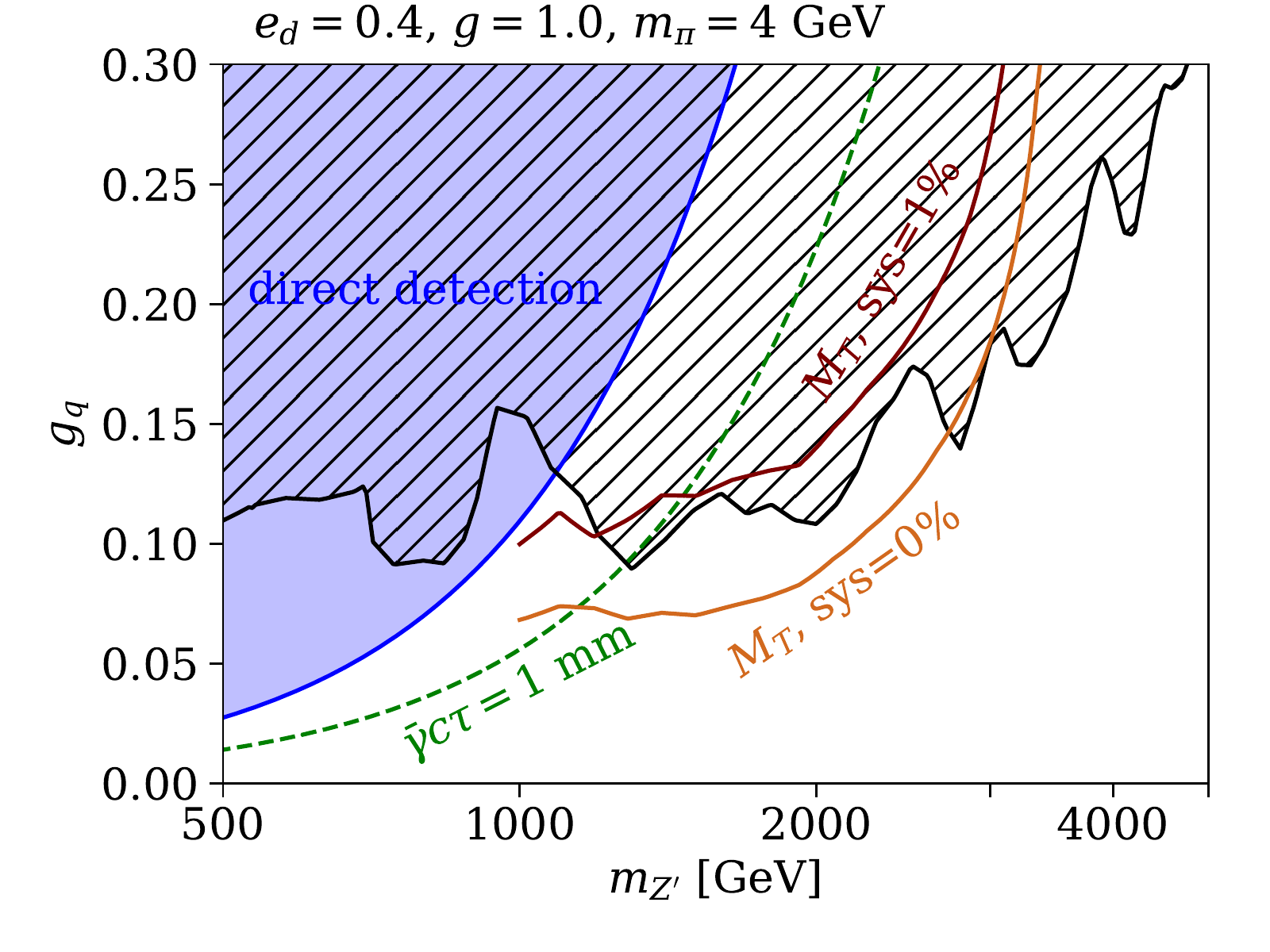}
	\includegraphics[width=0.495\columnwidth]{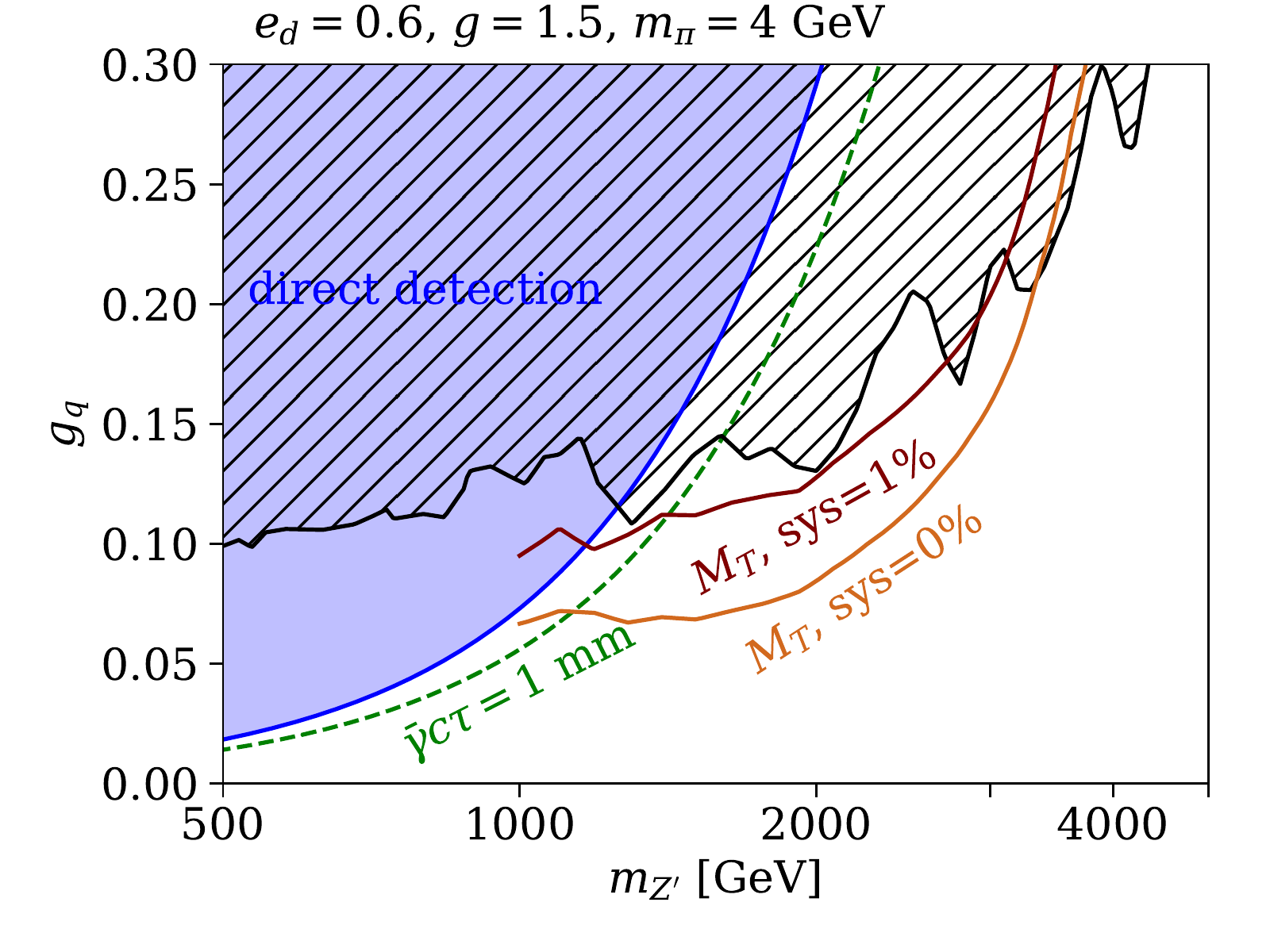}
	\caption{Expected sensitivity of the $M_T$-based analysis proposed in Ref.~\cite{Cohen:2017pzm} compared to the direct detection constraints from figure~\ref{fig:directdetection} and the LHC exclusion bounds from figure~\ref{fig:lhc_limits_etmiss} for different assumptions on systematic uncertainties. \label{fig:lhc_limits_mt}}
\end{figure}

We estimate the sensitivity of this search by calculating the likelihood ratio $L$ of signal+background and background-only and requiring $-2 \log L > 3.84$ in order for a given parameter point to be testable. The resulting sensitivity estimate is shown in figure~\ref{fig:lhc_limits_mt} in comparison to the combined bounds from existing searches (see figure~\ref{fig:lhc_limits_etmiss}). We find that the $M_T$ search has the power to probe larger $Z'$ masses than the missing energy searches discussed above and can be competitive with searches for resonances in visible decays for sufficiently large $e_\mathrm{d}$.\footnote{The projected sensitivity of the $M_T$ analysis extends into the parameter region where not all $\rho^0$ decays can be treated as prompt. Jets originating from displaced vertices would likely be rejected in such an analysis, which would reduce the sensitivity slightly compared to what is shown in figure~\ref{fig:lhc_limits_mt}.}

The $M_T$ search that we have implemented is not optimised for $m_{Z'} < 1\,\mathrm{TeV}$ and hence we do not show sensitivity projections in this mass range. Searches for low-mass resonances decaying into jets are notoriously difficult to observe because of large QCD backgrounds. While the missing energy requirement is expected to improve the situation, reliable estimates of backgrounds with fake missing energy due to misreconstructed jets are difficult to obtain. To suppress these backgrounds it may be necessary to require additional particles in the event, for example a high-$p_T$ photon from ISR (in analogy to the recent ATLAS search presented in Ref.~\cite{Aaboud:2019zxd}). We leave a detailed study of this mass range to future work.

We finally note that, following Ref.~\cite{Cohen:2017pzm}, the sensitivity estimate shown in figure~\ref{fig:lhc_limits_mt} assumes no systematic uncertainties. However, as shown in figure~\ref{fig:MTspectrum}, backgrounds are large and the signal is rather broad. Hence, even small uncertainties in the shape of the background may render a potential signal unobservable. We find that, in order to achieve a sensitivity close to the one shown in figure~\ref{fig:lhc_limits_mt}, systematic uncertainties must be smaller than 1\%. For the more realistic assumption of 5\% systematic uncertainties, the sensitivity of the $M_T$ offers no improvement over the searches discussed in section~\ref{sec:met_searches}.\footnote{We have also implemented an analysis in which the background shape is obtained from a fit to data, similar to what is commonly done when searching for a peak in $m_{jj}$ distributions. The sensitivity estimate obtained in this case is similar to the one for 5\% systematic uncertainties, i.e.\ it does not probe presently unexplored parameter space.} Detailed studies of systematic uncertainties will therefore be essential to obtain realistic sensitivity estimates.

\section{Conclusions}
\label{sec:conclusions}

In this work we have investigated how the phenomenology of strongly interacting dark sectors depends on their internal structure and on their interactions with the Standard Model. For simplicity, we have considered portal interactions arising from a $Z'$ mediator with vector couplings to both dark and visible quarks, but many of our results are insensitive to the details of this interaction. Below the confinement scale the dark quarks form $\pi$ and $\rho$ mesons, such that the dark sector can be characterised in terms of the masses $m_\pi$ and $m_\rho$ and the coupling strength $g$ of their interactions.

The case of two dark quarks with opposite $U(1)'$ charges turns out to be particularly interesting as all dark pions are then stable, allowing for a non-zero DM relic density as well as avoiding constraints from late decays of DM particles. The $\rho^0$ on the other hand has a sizeable decay width into SM particles, which is sufficient to establish thermal equilibrium between the dark sector and the visible sector in the early Universe. Dark sector freeze-out then proceeds via the forbidden annihilation process $\pi\pi \to \rho\rho$. The observed relic abundance can hence be reproduced independently of the strength of the portal interactions, provided $m_\pi$ and $m_\rho$ are sufficiently close. For example, for $m_\pi = 4 \, \mathrm{GeV}$ and $g = 1$ we require $m_\rho \approx 5 \, \mathrm{GeV}$. The observation that the cosmological properties of the dark sector and the relic abundance of dark mesons are largely independent of the interactions between the dark sector and the Standard Model justifies our simplified description. Additionally, the details of our dark sector have been chosen in a way to allow for a fully consistent cosmological picture to emerge, allowing us to consider LHC phenomenology from a dark sector that can realistically account for the observed DM in our universe.

Experimental predictions, on the other hand, depend more sensitively on the assumed portal interactions. For the $Z'$ mediator we consider, charged dark pions interact with SM nuclei via vector boson exchange, which gives rise to spin-independent scattering in direct detection experiments. By combining results from a number of different direct detection experiments, we obtain strong constraints on the coupling strength of the $Z'$. Based on these constraints we have identified dark pion masses of the order of a few GeV as particularly interesting. Even for dark pions in this mass range, direct detection constraints require the $Z'$ mediator to be either heavy or weakly coupled. In the present work, we have explored the former possibility, which leads to interesting implications for LHC physics.

LHC phenomenology for this model is dominated by the on-shell production of the mediator (possibly in association with SM particles) and its subsequent decays into either visible or dark quarks. While the former case leads to di-jet resonances, that can be easily reconstructed, the latter case gives rise to dark showers. We emphasise that for a strongly-interacting dark sector the branching ratio into dark quarks is enhanced by colour and flavour factors, making the latter signature particularly important. Since most of the mesons in the dark shower are stable, one obtains so-called semi-visible jets, in which only a small fraction of the initial energy of the dark quarks can be detected.

We find that various existing LHC searches for missing energy can be sensitive to such a scenario. Nevertheless, these searches are not optimised for the case of dark showers, where the missing energy tends to be aligned with one or more visible jets, which is difficult to disentangle from QCD backgrounds from mis-reconstructed jets. We therefore consider an alternative approach, in which the information from all visible final states is combined to calculate the transverse mass $M_T$ to achieve better discrimination between signal and background. The expected sensitivity of such a search depends however sensitively on the assumed systematic uncertainties for the background distribution. We find that systematic uncertainties as low as 1\% will be necessary in order to improve upon existing searches.

At the same time we identify a number of exciting directions for future research. First of all, a generic prediction of the model that we study is that the $\rho^0$ can be long-lived, which would give rise to displaced vertices at the LHC. The corresponding production cross sections can be quite large, and it is conceivable that thousands of such events have already gone unnoticed at ATLAS and CMS. Ongoing detector upgrades as well as new analysis strategies make these signatures a promising target for future LHC runs. But even for prompt decays there is room for substantial improvements. The jets obtained from our model differ quite substantially from ordinary QCD jets, for example because heavy quarks are absent in the shower. It will therefore be very interesting to study whether a neural network trained to identify dark showers can help to suppress QCD backgrounds.

\acknowledgments

We thank Alexander M{\"u}ck, Pedro Schwaller and Susanne Westhoff for discussions. This  work  is  funded  by  the  Deutsche Forschungsgemeinschaft (DFG) through the Collaborative Research Center TRR 257 ``Particle Physics Phenomenology after the Higgs Discovery'', the Emmy Noether Grant No.\ KA 4662/1-1 and the Research Unit FOR 2239 ``New Physics at the Large Hadron Collider''.

\appendix

\section{Full Lagrangian}
\label{app:lagrangian}

The Lagrangian of the underlying dark $SU(3)$ gauge theory coupled to a $Z^\prime$ vector mediator reads
\begin{align*}
\mathcal{L} = &-\frac{1}{4}F_{\mu\nu}^aF^{\mu\nu, a} + \overline{q}_\mathrm{d}i\slashed{D}q_\mathrm{d} - \overline{q}_\mathrm{d} M_q q_\mathrm{d} \\
&-\frac{1}{4}Z'_{\mu\nu}Z'^{\mu\nu} + \frac{1}{2}m_{Z'}^2Z'_\mu Z'^\mu - g_q \, Z^\prime_\mu \, \sum_{q_{\mathrm{SM}}}^{} \overline{q}_\mathrm{SM}\gamma^\mu q_\mathrm{SM} \; ,
\end{align*}
where $q_\mathrm{d} = (q_{\mathrm{d},1}, q_{\mathrm{d},2})$ and $M_q = \text{diag}(m_q, m_q)$. The dark quark covariant derivative has the form
\begin{align*}
D_\mu q_\mathrm{d} = \left(\partial_\mu + i g_\mathrm{d} A_{\mathrm{d}\mu} + i e_\mathrm{d} Z'_\mu Q\right) q_\mathrm{d} \; ,
\end{align*}
where $A_\mu$ denotes the dark gluon field and $Q = \mathrm{diag}(1,-1)$ is the dark quark charge matrix.

The chiral EFT Lagrangian (up to fourth order in the pion fields) is given by
\begin{align}
\mathcal{L}_\mathrm{EFT} \;=\; &\mathrm{Tr}\left(D_\mu\pi D^\mu\pi\right) -\frac{2}{3f_\pi^2}\mathrm{Tr}\left(\pi^2D_\mu\pi D^\mu\pi - \pi D_\mu\pi\pi D^\mu\pi\right) \nonumber \\
&+ m_\pi^2\mathrm{Tr}\left(\pi^2\right) + \frac{m_\pi^2}{3f_\pi^2}\mathrm{Tr}\left(\pi^4\right) + \mathcal{O}\left(\frac{\pi^6}{f_\pi^4}\right) \nonumber \\
&-\frac{1}{4}\mathrm{Tr}\left(V_{\mu\nu}V^{\mu\nu}\right) + m_\rho^2\mathrm{Tr}\left(V^2\right) -\frac{e_\mathrm{d}}{g} Z^\prime_{\mu\nu} \mathrm{Tr}\left(QV^{\mu\nu}\right)  \; .
\label{eq:LEFT}
\end{align}
Here $\pi$ denotes the pion matrix
\begin{align*}
\label{eq:pionmatrix}
\pi = \pi^a T^a = \frac{1}{\sqrt{2}}
\begin{pmatrix}
\frac{\pi^0}{\sqrt{2}} & \pi^+ \\
\pi^- & -\frac{\pi^0}{\sqrt{2}} \\
\end{pmatrix} \; ,
\end{align*}
with the $SU(2)$ generators $T^a=\frac{\sigma^a}{2}$.
In a similar fashion, we have introduced
\begin{align*}
V_\mu = V_\mu^aT^a = \frac{1}{\sqrt{2}}
\begin{pmatrix}
\frac{\rho^0_\mu}{\sqrt{2}} & \rho^+_\mu \\
\rho^-_\mu & -\frac{\rho^0_\mu}{\sqrt{2}}
\end{pmatrix}
\end{align*}
and
\begin{align*}
V_{\mu\nu} = \partial_\mu V_\nu - \partial_\nu V_\mu - ig\left[V_\mu,V_\nu\right] \; .
\end{align*}
The pion covariant derivative is given by
\begin{align*}
D_\mu \pi = \partial_\mu \pi + ig \left[\pi,V_\mu\right] + i e_\mathrm{d} \left[\pi,Q\right]Z^\prime_\mu \; .
\end{align*}

Eq.~(\ref{eq:LEFT}) gives rise to kinetic mixing between the $Z'$ and the $\rho^0$. To recover canonical kinetic terms, while at the same time keeping the mass term diagonal, the interaction eigenstates $\tilde{Z}'$ and $\tilde{\rho}_0$ can be expressed in terms of the physical fields $Z'_\mu$ and $\rho^0_\mu$ as
\begin{equation}
\begin{pmatrix}
 \tilde{Z}' \\ \tilde{\rho}_0
\end{pmatrix}
=
\begin{pmatrix}
 \sec \epsilon & \quad & \sin \epsilon \, \frac{m_\rho^2}{m_{Z'}} \\ - \tan \epsilon + \frac{1}{2} \sin 2 \epsilon \, \frac{m_\rho^2}{m_{Z'}^2} & \qquad  & 1 - \sin^2 \epsilon \, \frac{m_\rho^2}{m_{Z'}^2} 
 \end{pmatrix}
\begin{pmatrix}
 Z' \\ \rho^0
\end{pmatrix}
 \; , 
 \end{equation}
where $\epsilon = \text{arcsin} (2 \, e_\mathrm{d} / g)$ and we have only kept terms up to second order in $m_\rho / m_{Z'}$. Note that the diagonalisation can only be performed if the matrix of kinetic terms is positive definite, which requires $e_\mathrm{d} / g < 1/2$.

Because of the mixing, the $\rho^0$ obtains couplings to SM quarks, which can be written as
\begin{equation}
 \mathcal{L}_\text{EFT} \supset \frac{2 \, e_\mathrm{d} \, g_q}{g} \frac{m_\rho^2}{m_{Z'}^2} \rho^0_\mu \sum_{q_{\mathrm{SM}}}^{} \overline{q}_\mathrm{SM}\gamma^\mu q_\mathrm{SM} \; .
\end{equation}
At the same time, the mixing modifies the interactions between the $Z'$, the $\rho^0$ and dark pions. For example, the trilinear interactions become
\begin{equation}
 \mathcal{L}_\text{EFT} \subset \left(- 2 \, e_\mathrm{d} \sqrt{1 - \frac{4 \, e_\mathrm{d}^2}{g^2}} \frac{m_\rho^2}{m_{Z'}^2} Z'_\mu + g \, \rho^0_\mu\right) \left[\pi^+\left(\partial^\mu\pi^-\right)-\left(\partial^\mu\pi^+\right)\pi^-\right] \; .
\end{equation}
Interestingly, the mixing does not modify the interactions between the $\rho^0$ and $\pi^\pm$ but strongly suppresses the interactions between the $Z'$ and dark pions. Thus, for $m_\rho \ll m_{Z'}$ the $\rho^0$ replaces the $Z'$ as the primary mediator between the dark and the visible sector at low energies.

\section{Sensitivity of DarkSide-50 to low-mass dark matter}
\label{app:darkside}

In this appendix we take a closer look at the sensitivity of the DarkSide-50 experiment to low DM masses. Ref.~\cite{Agnes:2018ves} claims to provide the world-leading bound on the spin-independent DM-nucleon scattering cross section of DM particles with masses between 1.8 GeV and 5 GeV. To derive this bound Ref.~\cite{Agnes:2018ves} adopts the Bezrukov model~\cite{Bezrukov:2010qa} for the ionisation yield $Q_\mathrm{y}$ in liquid noble gases, which uses Lindhard theory~\cite{Lindhard:1961zz} to predict the electronic stopping power of liquid argon and a model of recombination proposed by Thomas and Imel~\cite{Thomas:1987zz}. This model is then fitted to calibration data to obtain the detector response for low recoil energies. However, as discussed in Ref.~\cite{Bezrukov:2010qa} the use of Lindhard theory for very low energies is highly doubtful and deviations from the simple model are expected. Ref.~\cite{Bezrukov:2010qa} proposes to vary the functional form of the electronic stopping power by multiplying it with a correction factor $F(v / v_0)$. It has been shown in Ref.~\cite{Frandsen:2013cna} that such variations can substantially affect the sensitivity of direct detection experiments to light DM particles. Indeed, we will demonstrate that the sensitivity of DarkSide-50 relies heavily on the assumed extrapolation for $Q_\mathrm{y}$.

We implement the DarkSide-50 analysis as follows. The expected number of electrons produced at the interaction point is calculated via $\langle N_e \rangle = Q_\mathrm{y}(E_\mathrm{R}) \, E_\mathrm{R}$. We assume the fluctuations in $N_e$ to follow a Poisson distribution convoluted with a Gaussian distribution that accounts for the detector resolution. The width of the Gaussian distribution is determined as $\sigma_{N_e} = 0.33$ by fitting the detector response for $N_e = 2$.\footnote{This approach neglects a possible dependence of $\sigma_{N_e}$ on $N_e$, which cannot be inferred from the available information.} An overall detector acceptance of 0.43 is already accounted in the total exposure of $6786\,\mathrm{kg\,days}$.

We then calculate the predicted number of events in three different $N_e$-bins: $[4, 7]$, $[7, 10]$, $[10, 22]$. In the first two bins, one finds a clear discrepancy between the background model (predicting $\sim390$ and $\sim550$ events, respectively) and observation ($\sim680$ and $\sim630$ events), which suggests an unknown background component. In these bins we therefore do not perform any background subtraction and only include them in the total likelihood if the predicted number of DM events exceeds the number of observed events.
In the third bin, on the other hand, background prediction ($\sim 4080$ events) and observation ($\sim 4130$ events) are in good agreement, so that one can use the standard Poisson likelihood to perform a background subtraction.

\begin{figure}[t]
\centering
\includegraphics[width=0.7\textwidth]{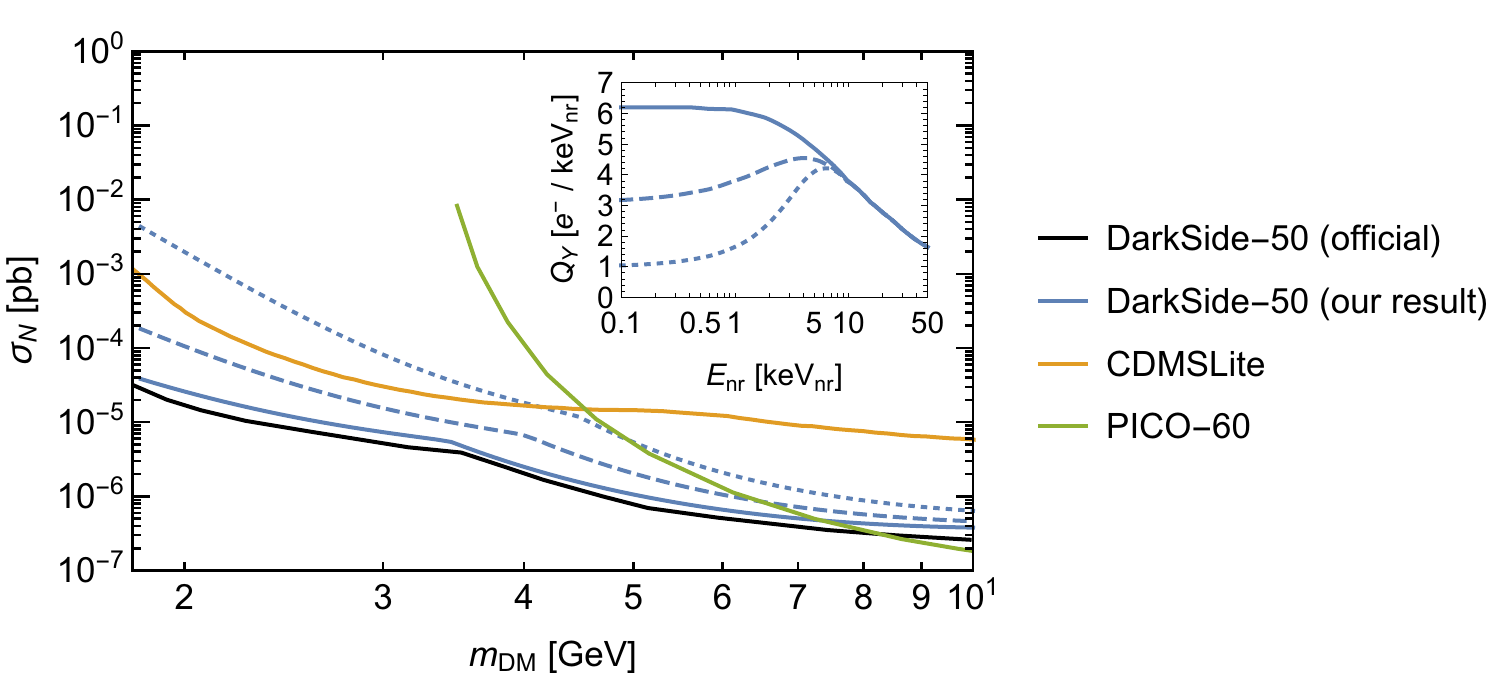}
\caption{Direct detection constraints on low-mass DM. For our reanalysis of DarkSide-50 we consider variations in the functional form of the ionisation yield $Q_\mathrm{y}$ as shown in the inset.}\label{fig:DarkSide}
\end{figure}

In spite of the rather crude implementation our analysis approximately reproduces the exclusion limit published by DarkSide-50 when making the same assumption on the ionisation yield $Q_\mathrm{y}$. We can therefore use our implementation to study the impact of varying $Q_\mathrm{y}$ on the exclusion limit. The impact of such variations are shown in figure~\ref{fig:DarkSide}. The solid black line corresponds to the official DarkSide-50 bound, the various blue lines correspond to the bounds obtained when adopting different functional forms for the ionisation yield $Q_\mathrm{y}$ as illustrated in the inset. We emphasise that all the variations that we consider differ only for $E_\mathrm{R} \lesssim 5\,\mathrm{keV}$, i.e.\ in an energy range for which no direct measurements of the ionisation yield are available.\footnote{The ionisation yield in this energy range may still be constrained using calibration data from DarkSide-50. However, to obtain robust constraints at very low recoil energies, the fit to calibration data should include variations in the functional form of $Q_\mathrm{y}$ like the ones considered here.}

Figure~\ref{fig:DarkSide} clearly illustrates the strong dependence of the claimed exclusion limit on the assumed functional form of the ionisation yield used for the extrapolation to lower recoil energies. Indeed, the bound from DarkSide-50 for DM masses of a few GeV can be suppressed by more than an order of magnitude, such that the strongest constraints in this mass range arise from CDMSLite~\cite{Agnese:2017jvy} and PICO-60~\cite{Amole:2017dex} (also shown in figure~\ref{fig:DarkSide}). We conclude that the DarkSide-50 exclusion limit is not robust unless theoretical calculations or experimental measurements can provide further evidence in support of the assumed functional form of $Q_\mathrm{y}$. We therefore do not include the exclusion limit from DarkSide-50 in our analysis.

\bibliographystyle{JHEP_improved}
\bibliography{simp}

\end{document}